\documentclass[usenatbib]{mn2e}
\usepackage{psfig}

\usepackage{epsfig}
\usepackage{amsmath}

\def\apj{ApJ}
\def\apjl{ApJL}
\def\mnras{MNRAS}

\def\araa{ARAA}
\def\aap{A\&A}
\def\aj{AJ}
\def\apjs{ApJS}

\def\nat{Nature}

\def\gs{\mathrel{\raise0.35ex\hbox{$\scriptstyle >$}\kern-0.6em\lower0.40ex\hbox{{$\scriptstyle \sim$}}}} 
\def\ls{\mathrel{\raise0.35ex\hbox{$\scriptstyle <$}\kern-0.6em\lower0.40ex\hbox{{$\scriptstyle \sim$}}}}

\def\Wm2{\,\hbox{W}\,\hbox{m}^{-2}} 
\def\gsim{\mathrel{\raise0.35ex\hbox{$\scriptstyle >$}\kern-0.6em\lower0.40ex\hbox{{$\scriptstyle \sim$}}}} 
\def\lsim{\mathrel{\raise0.35ex\hbox{$\scriptstyle <$}\kern-0.6em\lower0.40ex\hbox{{$\scriptstyle \sim$}}}}

\begin{document}

\title[ALMA {\rm [C{\sc ii}]} detections of SMGs in the ECDFS]{An ALMA
  survey of Sub-millimetre Galaxies in the Extended Chandra Deep Field
  South: Detection of [C{\sc ii}] at \emph{z}\,=\,4.4}

\author[Swinbank et al.]
{\parbox[h]{\textwidth}
{A.\,M.\ Swinbank,$^{\,1,*}$ 
A.\ Karim,$^{\,1}$ 
Ian Smail,$^{\,1}$ 
J.\ Hodge,$^{\,2}$ 
F.\, Walter,$^{\,2}$\\ 
F.\ Bertoldi,$^{3}$
A.\,D.\ Biggs,$^{\,4}$ 
C.\, de Breuck,$^{\,4}$ 
S.\,C.\ Chapman,$^{\,5}$ 
K.\,E.\,K\ Coppin,$^{\,6}$ 
P.\, Cox,$^{\,7}$ 
A.\,L.\,R.\, Danielson,$^{\,1}$ 
H.\ Dannerbauer,$^{8}$
R.\,J.\ Ivison,$^{\,9,10}$ 
T.\,R.\ Greve,$^{\,11}$ 
K.\, K.\ Knudsen,$^{\,12}$ 
K.\,M.\ Menten,$^{\,3}$ 
J.\ M.\, Simpson,$^{\,1}$ 
E.\ Schinnerer,$^{\,2}$ 
J.\,L.\ Wardlow,$^{13}$
A.\ Wei\ss,$^{\,14}$ 
P.\ van der Werf$^{\,15}$ 
}
\vspace*{6pt} \\ 
$^1$Institute for Computational Cosmology, Durham University, South Road, Durham, DH1 3LE, UK\\
$^2$Max-Planck-Institut f\"ur Astronomie, K\"onigstuhl 17, D-69117 Heidelberg, Germany\\
$^3$Argelander-Instutute of Astronomy, Bonn University, Auf dem Huegel 71, D-53121 Bonn, Germany \\
$^4$European Southern Observatory, Karl-Schwarzschild Strasse 2, D-85748 Garching, Germany\\
$^5$Institute of Astronomy, University of Cambridge, Madingley Road, Cambridge CB3 0HA\\
$^6$Department of Physics, McGill University, 3600 Rue University, Montréal, QC H3A 2T8, Canada\\
$^7$Institut de Radioastronomie Millim\`etrique, 300 rue de la piscine, F-38406 Saint-Martin d'H\`eres, France\\
$^8$Universit\"at Wien, Institut f\"ur Astrophysik,  T\"urkenschanzstra\ss e 17, 1180 Wien, Austria\\
$^9$UK Astronomy Technology Centre, Science and Technology Facilities Council, Royal Observatory, Blackford Hill, Edinburgh EH9 3HJ\\
$^{10}$Institute for Astronomy, University of Edinburgh, Blackford Hill, Edinburgh EH9 3HJ\\
$^{11}$University College London, Department of Physics \& Astronomy, Gower Street, London, WC1E 6BT, UK \\
$^{12}$Department of Earth and Space Sciences, Chalmers University of Technology, Onsala Space Observatory, SE-43992 Onsala, Sweden\\
$^{13}$Department of Physics \& Astronomy, University of California, Irvine, CA 92697, USA\\
$^{14}$Leiden Observatory, Leiden University, PO Box 9513, 2300 RA Leiden, Netherlands\\
$^{*}$Email: a.m.swinbank@dur.ac.uk\\
}

\maketitle

\begin{abstract}
  We present ALMA 870-$\mu$m (345\,GHz) observations of two
  sub-millimetre galaxies (SMGs) drawn from an ALMA study of the 126
  sub-millimeter sources from the LABOCA Extended {\it Chandra} Deep
  Field South Survey (LESS).  The ALMA data identify the counterparts
  to these previously unidentified sub-millimeter sources and
  serendipitously detect bright emission lines in their spectra which
  we show are most likely to be [C{\sc ii}]\,157.74\,$\mu$m emission
  yielding redshifts of $z$\,=\,4.42 and $z$\,=\,4.44.  This blind
  detection rate within the 7.5-GHz bandpass of ALMA is consistent with
  the previously derived photometric redshift distribution of SMGs and
  suggests a modest, but not dominant ($\lsim $\,25\%), tail of
  870-$\mu$m selected SMGs at $z\gsim$\,4.  We find that the ratio of
  $L_{\rm [CII]}$\,/\,$L_{\rm FIR}$ in these SMGs is much higher than
  seen for similarly far-infrared-luminous galaxies at $z\sim$\,0,
  which is attributed to the more extended gas reservoirs in these
  high-redshift ULIRGs.  Indeed, in one system we show that the [C{\sc
      ii}] emission shows hints of extended emission on $\gsim$\,3\,kpc
  scales.  Finally, we use the volume probed by our ALMA survey to show
  that the bright end of the [C{\sc ii}] luminosity function evolves
  strongly between $z$\,=\,0 and $z\sim$\,4.4, reflecting the increased
  ISM cooling in galaxies as a result of their higher star-formation
  rates.  These observations demonstrate that even with short
  integrations, ALMA is able to detect the dominant fine structure
  cooling lines from high-redshift ULIRGs, measure their energetics,
  spatially resolved properties and trace their evolution with
  redshift.
\end{abstract}

\begin{keywords}
  galaxies: starburst, galaxies: evolution, galaxies: high-redshift
\end{keywords}


%
%
%
\section{Introduction}

A significant fraction of the obscured star formation at $z\gg 1$
arises from the most luminous galaxies
\citep[e.g.][]{Blain99b,LeFloch09}: ultra-luminous infrared galaxies
(ULIRGs; \citealt{Sanders96}) with bolometric luminosities of
$\gsim$10$^{12-13}$\,L$_\odot$ and star-formation rates
$\gsim$\,100--1000\,M$_\odot$\,yr$^{-1}$.  At high redshift, these
galaxies are some of the brightest sources in the sub-millimetre
wave-band and so are frequently called ``sub-millimetre galaxies''
(SMGs).

The spectral energy distribution (SED) of the dust emission in these
luminous dusty galaxies has a Rayleigh-Jeans tail that produces a
negative K-correction in the sub-millimetre wave-band, yielding a near
constant flux density--luminosity dependence with redshift over the range
$z\sim $\,1--6 \citep{Blain02}.  Although the volume density of the SMG
population evolves rapidly with redshift, (showing a 1000-fold increase
in the space density over the $\sim$\,10\,Gyrs to $z\sim$\,2.5;
\citealt{Chapman05a}), the volume density appears to subsequently
decline above $z\gsim$\,3.  This decline is not a result of a reduction
in their apparent flux density in the sub-millimetre, but a real reduction in
the volume density of SMGs \citep{Wardlow11}.

However, determining the precise strength of the decline in the SMG
volume density requires locating the counterparts of high-redshift
SMGs.  This is problematic due to the poor resolution of single dish
sub-millimetre maps, which means that SMGs have to be identified
through correlations between their sub-millimetre emission and that in
other wave-bands where higher spatial resolution is available (usually
the radio and/or mid-infrared
e.g.\ \citealt{Ivison02,Chapman04b,Ivison07,Pope06}).  These
identifications are probabilistic as they rely on empirical
correlations that have both significant scatter and which may evolve
with redshift \citep[e.g.\ ][]{CarilliYun00}.  Moreover, the SEDs in
these other wave-bands have positive K-corrections and hence can miss
the highest redshift counterparts.  Indeed in sub-millimetre surveys
typically 30--50\% of SMGs lack ``robust'' counterparts in the radio or
mid-infrared, and these may represent an unidentified tail of
high-redshift SMGs (e.g.\ \citealt{Biggs11}, see also
\citealt{Lindner11}).  To circumvent this problem requires identifying
SMGs using sub-/millimetre interferometers
\citep[e.g.\ ][]{Gear00,Frayer00}, but until recently their sensitivity
has been too low to locate large numbers of SMGs
\citep[e.g.][]{Dannerbauer02,Younger07,Wang11,Smolcic12}.  However,
with the commissioning of the Atacama Large Millimeter Array (ALMA), we
can now construct the large samples of precisely-located SMGs needed to
unambiguously study their properties and test galaxy formation models.

Whilst the bulk of the bolometric emission from SMGs is radiated
through continuum emission from dust grains in the rest-frame
far-infrared, superimposed on this are a series of narrow atomic and
molecular emission lines.  By far the strongest of these arise from the
atomic fine-structure transitions of Carbon, Nitrogen and Oxygen in the
far-infrared, along with weaker, but more commonly-studied molecular
lines visible at millimetre wavelengths (e.g.\ $^{12}$CO, HCN).
These bright emission lines are an important pathway by which the dense
gas in these galaxies cools, and so provide a unique tracers of the
star-formation process in these galaxies.

The brightest and best-studied of the atomic lines in the far-infrared
is the $^2P_{3/2}$--$^2P_{1/2}$ fine structure line of singly ionised
carbon at 157.74-$\mu$m (here-after [C{\sc ii}]).  Much of the [C{\sc
    ii}] emission from galaxies arises from the warm and dense
photo-disassociation regions (PDRs) that form on the UV-illuminated
surface of molecular clouds, although the [C{\sc ii}] flux from diffuse
H{\sc ii} regions or from cool, diffuse interstellar gas can also be
significant \citep[e.g.\ ][]{Madden93,Lord96}.  The [C{\sc ii}] emission
line therefore provides an indication of the gas content and the extent
of the gas reservoir in a galaxy.  Far-infrared surveys of low-redshift
galaxies from the Kuiper Airborne Observatory and {\it ISO} have shown
that the [C{\sc ii}] line can comprise $\gsim 1$\% of the total
bolometric luminosity \citep{Stacey91,Brauher08,Gracia-Carpio11}.  This
bright line is thus ideally suited for deriving redshifts for obscured
galaxies and investigating their dynamics and star-formation
properties.

Early searches for [C{\sc ii}], by necessity, focused on the
highest-redshift far-infrared sources, $z>$\,4, where the [C{\sc ii}]
line is shifted into the atmospheric windows in the sub-millimetre
\citep[e.g.\ ][]{Ivison98b,Maiolino05,Maiolino09,Wagg10}.  Most of
these sources host powerful AGN (as well as being ULIRGs) and it was
noted that their [C{\sc ii}] lines were weak relative to $L_{\rm FIR}$,
demonstrating the same behavior as seen in AGN-dominated ULIRGs in the
local Universe.  However, more recent observations of the [C{\sc ii}]
emission in high redshift star-formation dominated ULIRGs has shown
that the [C{\sc ii}] emission can be as bright as in local,
low-luminosity galaxies, $L_{\rm [CII]}$\,/\,$L_{\rm
  FIR}\sim$\,0.1--1\%
\citep[e.g.][]{SHD08,Ivison10eyelash,Stacey10,Valtchanov11}.  This has
been interpreted as due to the lower ionisation field arising from more
widely distributed star-formation activity within these systems, in
contrast to the compact nuclear star formation seen in low-redshift
ULIRGs \citep[e.g.][]{Sakamoto08}.  The strength of [C{\sc ii}] (and
other atomic lines), and its relative strength to the far-infrared
luminosity can therefore be used to probe the physical properties of
the interstellar medium in high-redshift galaxies.

We have recently undertaken an ALMA Cycle 0 study at 870$\mu$m
(345\,GHz) of the 126 sub-millimeter sources located in the
0.5$^{\circ}$\,$\times$\,0.5$^{\circ}$ LABOCA Extended {\it Chandra}
Deep Field South Survey (``LESS''; \citealt{Weiss09}), the most uniform
sub-millimetre survey of its kind to date.  These ALMA data yield
unambiguous identifications for a large fraction of the sub-millimeter
sources, directly pin-pointing the SMG responsible for the
sub-millimetre emission to within $<$\,0.2$''$ (Hodge et al.\ 2012, in
prep), without recourse to statistical radio/mid-infrared associations.
In this letter, we present ALMA observations of two of the SMGs from
our survey for which we are able to derive their redshifts from
serendipitous identification of the [C{\sc ii}] emission line in the
ALMA data-cubes.  We use the data to measure the energetics of the
dominant fine structure lines, search for spatially resolved velocity
structure and to provide an estimate of the evolution of the [C{\sc
    ii}] luminosity function with redshift.  We adopt a cosmology with
$\Omega_{\Lambda} =$\,0.73, $\Omega_{m}=$\,0.27, and $H_{\rm
  0}=$\,72\,km\,s$^{-1}$\,Mpc$^{-1}$ in which 1$''$ corresponds to a
physical scale of 6.7\,kpc at $z\sim$\,4.4.

%
%
\begin{figure*}
  \centerline{\psfig{file=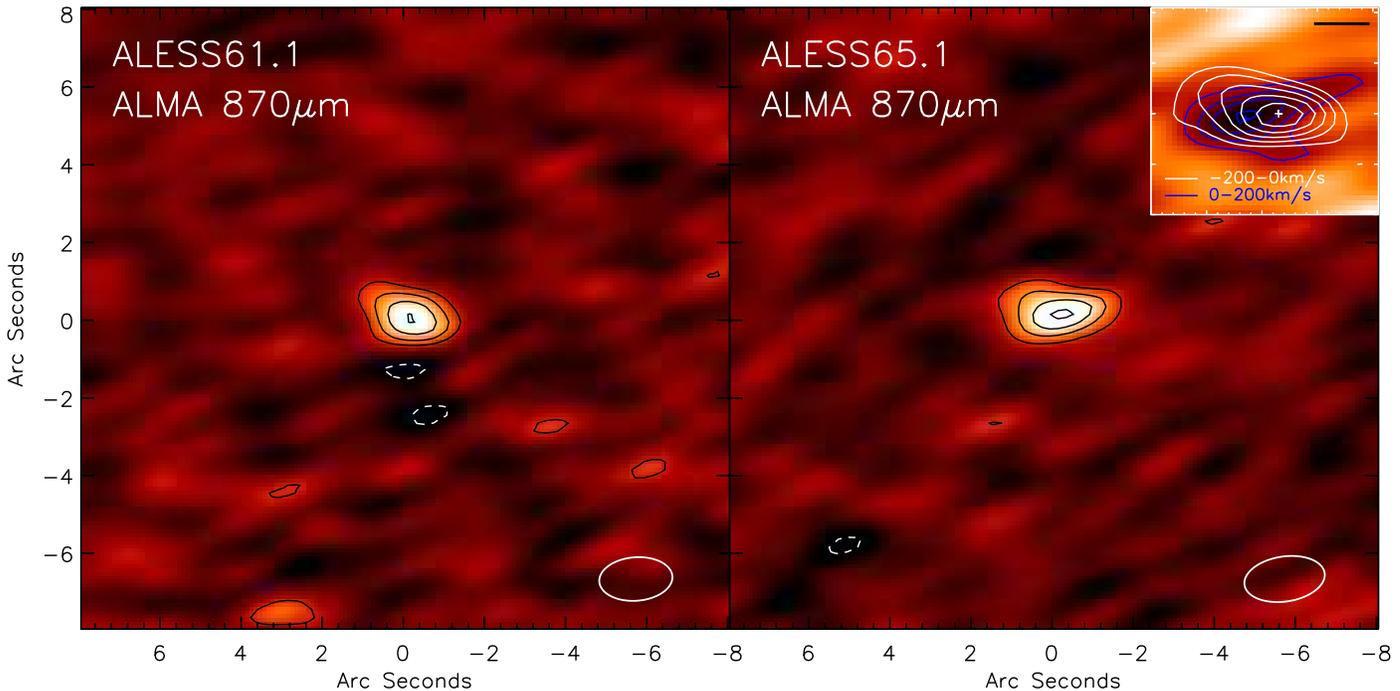,angle=0,width=7in}}
\caption{{\it Top:} 870$\mu$m (345\,GHz) velocity integrated ALMA maps
  of LESS\,61 and LESS\,65 from the Band 7 ALMA observations in compact
  configuration (these maps include the continuum and emission lines
  and are naturally weighted).  In both cases, we detect strong
  870$\mu$m emission in both sub-millimeter sources associated with a
  single SMG.  The r.m.s. noise in these maps is 0.44 and
  0.42\,mJy\,beam$^{-1}$ for LESS\,61 and LESS\,65 respectively.
  Contours denote $\pm$\,3, 5, 7...$\sigma$.  In each panel we also
  show the 50\% contour of the synthesised beam which is approximately
  1.8$''\times$\,1.2$''$ in both cases.  These maps are centered at
  $\alpha:$\,03\,32\,45.87 $\delta:$ $-$28\,00\,23.3 and
  $\alpha:$\,03\,32\,52.26 $\delta:$ $-$27\,35\,26.3 for LESS\,61 and
  LESS\,65 respectively (Table~1).  The inset in LESS\,65 shows the
  velocity structure in the [C{\sc ii}] in this source.  The colour
  scale shows the (continuum subtracted) velocity integrated cube
  between $-$210--0\,km\,s$^{-1}$ with the blue contours at 2, 3, 4,
  5...$\sigma$.  The white contours denote the velocity integrated
  emission 0--210\,km\,s$^{-1}$ (also at 2, 3, 4, 5...$\sigma$). The
  spatial offset between the blue-shifted and red-shifted emission in
  these two images is 0.50\,$\pm$\,0.25\,$''$ (3.3\,$\pm$\,1.7\,kpc).
  If this spatial and velocity offsets represents rotating gas
  reservoir, then we estimate a dynamical mass of $M_{\rm
    dyn}\sim$\,3.5\,$\times$\,10\,$^{10}$\,$\sin^2(i)$\,M$_{\odot}$.
  In this inset, the solid bar shows a scale of 1$''$. }
\label{fig:maps}
\end{figure*}

%
%
\section{Observations and Reduction}

Observations of the 126 sub-millimeter sources in the LESS survey were
obtained with ALMA at 345\,GHz (Band 7) with a dual polarisation setup
in the compact configuration, yielding a synthesised beam of
$\sim$\,1.8$''\times$1.2$''$.  Our observations cover 7.5\,GHz
band-width, split between the upper- and lower-sidebands,
336.1--339.8\,GHz and 348.1--351.9\,GHz.  The observations employed 15
antennae and data for these two sources were obtained between 2011
October and 2011 November in good conditions, PWV\,$\lsim$\,0.5\,mm.
The primary beam of the ALMA dishes, $\sim$\,18$''$ FWHM at our
observing frequency, is sufficient to encompass the error-circles of
the SMGs from the LESS maps, $\lsim$\,5$''$ \citep{Weiss09}, even in
confused situations.  The observing frequency was selected to match the
original LABOCA discovery map to ensure that the ALMA and LABOCA flux
densities could be easily compared.  Each galaxy was observed for a
total of $\sim$\,120\,seconds, with phase, bandpass and flux
calibration based on J0403-360, J0538-440 and Mars, respectively.  The
data were processed with the Common Astronomy Software Application
(CASA; \citealt{McMullin07}), and we constructed both velocity
integrated maps and datacubes with 62.5\,MHz (54.5\,km\,s$^{-1}$)
binning.  The velocity integrated continuum maps for the two sources
discussed here reach noise levels of $\sigma$\,=\,0.44 and
0.42\,mJy\,beam$^{-1}$ for LESS\,61 and LESS\,65 respectively.  This is
a factor $\sim$\,3\,$\times$ more sensitive than the original LABOCA
discovery map, while more critically the beam is $\sim$150\,$\times$
smaller in area than that of LABOCA (with a corresponding reduction in
the positional uncertainty of detected sources).  In the datacubes, the
r.m.s. sensitivity is $\sim$\,3.8\,mJy\,beam$^{-1}$ per
54.5\,km\,s$^{-1}$ (62.5\,MHz) channel.  The full catalog of the ALMA
SMGs will be published in Hodge et al.\ (2012, in prep), whilst the
ALMA 870$\mu$m counts and multi-wavelength properties of the ALMA SMGs
will be published in Karim et al.\ (2012, in prep) and Simpson et
al.\ (2012, in prep).

%
%
\begin{table}
\begin{center}
\caption{Co-ordinates and Photometry}
\begin{tabular}{lcc}
\hline
\hline
                         & ~~~ALESS\,61.1                     &  ~~~ALESS\,65.1             \\
\hline
ID$^{a}$                 & J033245.6$-$280025        &  J033252.4$-$273527 \\  
R.A.\ (ALMA)$^b$          & 03\,32\,45.87                  & 03\,32\,52.26         \\
Dec.\ (ALMA)$^b$          & $-$28\,00\,23.3                & $-$27\,35\,26.3         \\
S$_{\rm 870}$ (LABOCA)     & 5.8\,$\pm$\,1.2                   & 5.9\,$\pm$\,1.2           \\
S$_{\rm 870}$ (ALMA)$^{c}$ & 4.32\,$\pm$\,0.44                  & 4.24\,$\pm$\,0.49          \\
0.365$\mu$m              & $<$\,1.2\,$\times$\,10$^{-4}$       & $<$\,9.9\,$\times$\,10$^{-5}$ \\
0.350$\mu$m              & $<$\,9.9\,$\times$\,10$^{-5}$       & $<$\,9.9\,$\times$\,10$^{-5}$ \\
0.460$\mu$m              & $<$\,5.7\,$\times$\,10$^{-5}$       & $<$\,1.8\,$\times$\,10$^{-4}$ \\
0.538$\mu$m              & $<$\,6.9\,$\times$\,10$^{-5}$       & $<$\,3.6\,$\times$\,10$^{-4}$ \\
0.651$\mu$m              & $<$\,6.3\,$\times$\,10$^{-5}$       & $<$\,6.3\,$\times$\,10$^{-5}$ \\
0.833$\mu$m              & 4.7\,$\pm$\,1.1\,$\times$\,10$^{-4}$  & $<$\,1.2$\times$\,10$^{-3}$ \\
0.850$\mu$m              & 8.5\,$\pm$\,1.0\,$\times$\,10$^{-4}$  & $<$\,3.6$\times$\,10$^{-4}$ \\
0.903$\mu$m              & 7.3\,$\pm$\,1.6\,$\times$\,10$^{-4}$  & $<$\,2.4$\times$\,10$^{-3}$ \\
2.2$\mu$m                & 3.0\,$\pm$\,0.3\,$\times$\,10$^{-3}$  & $<$\,2.7$\times$\,10$^{-3}$ \\
3.6$\mu$m                & 3.3\,$\pm$\,0.3\,$\times$\,10$^{-3}$  & 2.1\,$\pm$\,0.2\,$\times$\,10$^{-3}$ \\
4.5$\mu$m                & 3.9\,$\pm$\,0.2\,$\times$\,10$^{-3}$  & 1.4\,$\pm$\,0.2\,$\times$\,10$^{-3}$ \\
5.8$\mu$m                & 2.7\,$\pm$\,1.0\,$\times$\,10$^{-3}$  & 1.4\,$\pm$\,1.0\,$\times$\,10$^{-3}$ \\
8.0$\mu$m                & 5.4\,$\pm$\,0.7\,$\times$\,10$^{-3}$  & 3.0\,$\pm$\,0.8\,$\times$\,10$^{-3}$ \\
24$\mu$m                 & 0.036\,$\pm$\,0.002              & $<$\,0.006                      \\
250$\mu$m                & 4.3\,$\pm$\,1.5	            & $<$\,2.7                        \\
350$\mu$m                & 7.4\,$\pm$\,1.6                  & 7.6\,$\pm$\,1.4                 \\
500$\mu$m                & 10.2\,$\pm$\,1.7                 & 10.2\,$\pm$\,1.5                \\
1.4\,GHz                 & $<$\,0.025                      & $<$\,0.025                       \\
\hline
\label{table:table1}
\end{tabular}
\end{center}
\noindent{\footnotesize Notes: All flux densities are in mJy and all
  limits are 3$\sigma$.\\ $^{a}$ LESS ID from the LABOCA catalog in
  \citet{Weiss09}.  $^b$ Co-ordinates in J2000.  $^{c}$ The ALMA flux
  densities have been primary beam corrected (the primary beam
  corrections are a factor 1.12 and 1.04 for ALESS\,61.1 and
  ALESS\,65.1 respectively).  The $BVRIzK$+IRAC photometry comprises
  VIMOS ($U$), MUSYC ($BVRIz$), \emph{HST} ($z_{\rm 850LP}$); HAWK-I
  ($K$) and IRAC\,3.6--8$\mu$m imaging \citep[see][for
    details]{Wardlow11}.  The \emph{Herschel} SPIRE photometry is
  measured using archival imaging and the photometry has been
  de-blended for nearby sources.  The radio flux density limits are
  taken from the VLA 1.4\,GHz imaging used in \citet{Biggs11}.  \hfil }
\end{table}

%
%
\begin{figure}
  \centerline{\psfig{file=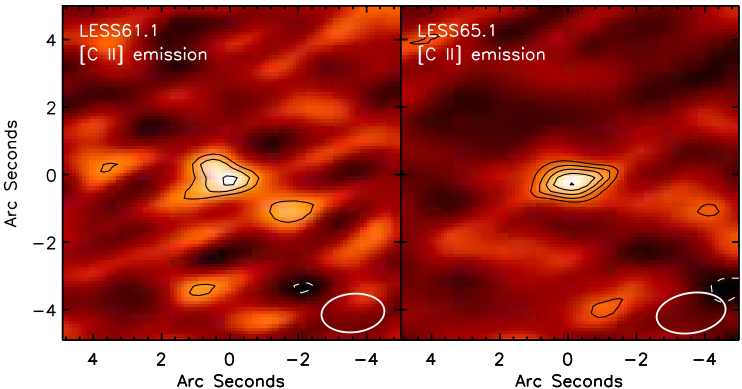,angle=90,width=3.8in}}
\caption{Continuum subtracted [C{\sc ii}] emission line maps of
  LESS\,61.1 and LESS\,65.1.  These are generated by subtracting the
  continuum and then integrating the cube between
  $\pm$\,2\,$\times$\,FWHM of the emission line.  In all panels the
  contours denote S\,/\,N levels denote $\pm$\,3, 4, 5...$\sigma$.}
\label{fig:maps2}
\end{figure}

%
%
\section{Analysis and Discussion}

\citet{Wardlow11} derive a photometric redshift distribution for the
statistically identified radio- and mid-infrared counterparts of the
LESS sub-millimeter sources, deriving a median $z$\,=\,2.5\,$\pm$\,0.5.
Taking the errors on the redshifts into account, we expect
3.5\,$\pm$\,0.5 SMGs should have far-infrared fine structure lines
(such as [C{\sc ii]}\,157.74$\mu$m, [O{\sc i}]\,145$\mu$m, [N{\sc
    ii}]\,122$\mu$m or [O{\sc iii}]\,88\,$\mu$m) within the 7.5\,GHz
bandwidth of ALMA.  By far the brightest of these lines in local ULIRGs
is the [C{\sc ii}] emission line, with a median equivalent width
$\gsim$\,10\,$\times$ brighter than any of the other lines
\citep{Brauher08}.  Focusing just on [C{\sc ii}], we expect 1.5$\pm$0.5
SMGs in our sample will have detectable [C{\sc ii}] emission in the
ALMA band-pass, with corresponding redshift ranges of
$z$\,=\,4.399--4.461 and $z$\,=\,4.590--4.656 for the upper and
lower-side bands.

As an initial step in our analysis of the ALMA maps of the LESS
sub-millimeter sources, we exploit the frequency coverage of our
observations to search for emission lines in the datacubes.  For each
$>$\,4-$\sigma$ SMG in the velocity-integrated ALMA maps within the
primary beam, we extract the spectra and search for emission lines by
attempting to fit a Gaussian emission line profile to the spectrum,
only accepting the fit if the $\Delta\chi^2$ provides a significant
improvement ($\Delta\chi^2>$\,25, or $\sim$\,5$\sigma$) over a
continuum-only fit.

\subsection{Source Identification}

\begin{figure}
  \centerline{
    \psfig{file=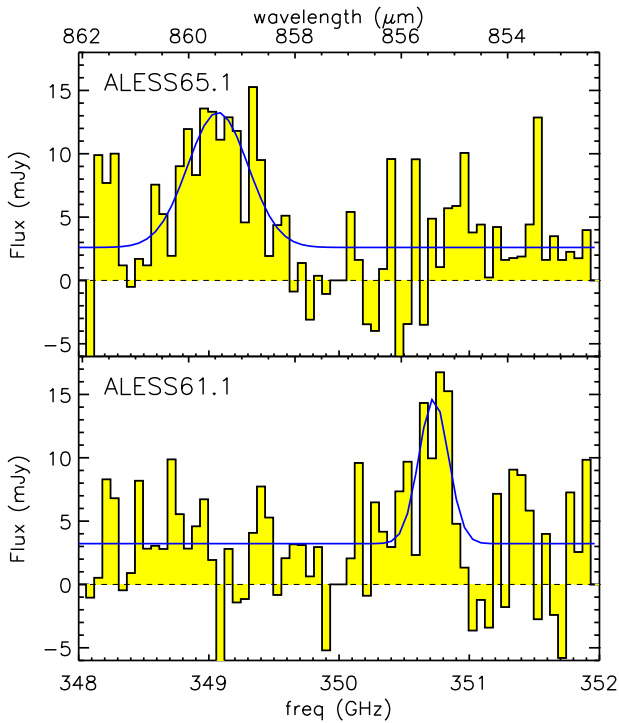,width=3.7in,angle=90}}
\caption{ALMA spectra of ALESS\,61.1 and ALESS\,65.1, extracted at the
  position of the peak sub-millimetre emission from the SMGs.  From the
  spectra, we identify bright emission lines in the upper side-band
  which we attribute to [C{\sc ii}] emission at $z$\,=\,4.419 and
  $z$\,=\,4.445 for ALESS\,61.1 and ALESS\,65.1, respectively.  The
  rest-frame equivalent widths of the lines are 0.44\,$\pm$\,0.06$\mu$m
  and 1.12\,$\pm$\,0.15\,$\mu$m for ALESS\,61.1 and ALESS\,65.1
  respectively, similar to the median equivalent width of [C{\sc ii}]
  in local LIRGs and ULIRGs (W$_{\rm
    o,[CII]}$\,=\,0.76\,$\pm$\,0.06\,$\mu$m; \citealt{Brauher08}).  In
  both cases, due to their selection both SMGs have significant
  continuum emission.}
\label{fig:spectra}
\end{figure}

In two sub-millimeter sources, LESSJ033245.6$-$280025 and
LESSJ033252.4$-$273527 (hereafter ALESS\,61.1 and ALESS\,65.1,
respectively following the notation in Hodge et al.\ 2012, in prep), we
identify bright emission lines with FWHM 230\,$\pm$\,25 and
490\,$\pm$\,35\,km\,s$^{-1}$ respectively in the spectra of these SMGs.
The significance of the emission lines (integrated over the line) is
S\,/\,N\,$\sim$\,5.3 and 7.0 for ALESS\,61.1 and ALESS\,65.1
respectively.  In Fig.~\ref{fig:maps} we show the velocity-integrated
(continuum plus line) ALMA maps of ALESS\,61.1 and ALESS\,65.1 while
the spectra are shown in Fig.~\ref{fig:spectra}.  To highlight the
  emission line detections in both ALMA SMGs, in Fig.~\ref{fig:maps2}
  we also show the continuum subtracted emission line maps.  In this
  figure, the map is made by integrating the continuum subtracted cube
  over $\pm$\,2\,$\times$\,FWHM of the emission line.

In both cases, the LABOCA 870$\mu$m source is identified with a single
high signal-to-noise (S\,/\,N\,$\sim$\,8) ALMA SMG in the velocity
integrated maps that is located to within $<$\,0.2$''$ (after
centroiding) and 3--4$''$ from their nominal LABOCA positions
\citep{Weiss09}.  \citet{Biggs11} could not identify robust or
tentative counterparts for either of these sub-millimeter sources
\citep[see also][]{Wardlow11} as no SMGs were detected at 1.4\,GHz
(3$\sigma\leq$\,25\,$\mu$Jy at 1.4\,GHz).  Both sources are very faint
in the \emph{Herschel}/SPIRE imaging at 250, 350, and 500\,$\mu$m (the
SMGs are weakly detected after de-blending nearby sources using the
24$\mu$m and radio as priors; Table~1).  The ALMA maps reveal that
ALESS\,61.1 is associated with a weak 24\,$\mu$m source that is also
visible in the optical--mid-infrared, but ALESS\,65.1 does not have any
counterparts in the optical or near-infrared and is only detected
weakly in the {\it Spitzer} IRAC imaging.  Together, this suggest that
these galaxies lie at $z\gsim$\,3.5 \citep{Wardlow11}.  With the
current data, we find no evidence for strong gravitational lensing of
either of these galaxies.

%
%

\subsection{Line Identification}

If the emission lines we identified are indeed [C{\sc ii}], then our
data indicate redshifts of $z$\,=\,4.419 and $z$\,=\,4.445 for
ALESS\,61.1 and ALESS\,65.1 respectively.  The rest-frame equivalent
widths of the emission lines are 0.44\,$\pm$\,0.06$\mu$m and
1.12\,$\pm$\,0.15\,$\mu$m, similar to the median equivalent width of
[C{\sc ii}] in local LIRGs and ULIRGs ($W_{\rm
  o,[CII]}$\,=\,0.76\,$\pm$\,0.06$\mu$m; \citealt{Brauher08}).  We note
that in the 7.5\,GHz spectral coverage at 345\,GHz, the emission lines
contribute 10--40\% of the continuum emission in the velocity
integrated cubes, as predicted by \citet{Smail11}.

We caution that there are other possible identifications for these
lines.  In particular, the emission could also be [O{\sc i}]\,145$\mu$m
($z\sim$\,4.9), [N{\sc ii}]\,122$\mu$m ($z\sim6.0$), [N{\sc
    ii}]\,205$\mu$m ($z\sim3.1$) or much higher redshift, [O{\sc
    iii}]\,88\,$\mu$m ($z\sim8.1$).  However, the luminosities of these
fine structure lines are expected to be a factor $\gsim$\,10\,$\times$
fainter than the [C{\sc ii}] emission
(e.g.\ \citealt{Brauher08,Walter09b,Nagao12,Decarli12}).  

Alternatively, the emission line could also be high-$J$ $^{12}$CO at
lower redshift (i.e. $z$\,=\,1.3, 1.6, 2.0 and 2.3 for $J$\,=\,7, 8, 9,
10 respectively).  However, the implied emission line luminosities
(L\,=\,1--7\,$\times$\,10$^{8}$\,L$_{\odot}$) are a factor
$\sim$\,10\,$\times$ higher than the $^{12}$CO line luminosities of any
local or high-redshift starbursts or AGN.  For example, Mrk\,231 has a
maximum $^{12}$CO luminosity (across all $^{12}$CO lines) of
$\lsim$\,3\,$\times$\,10$^7$\,L$_{\odot}$ \citep{vanderWerf10}, whilst
APM\,08279, a well studied high-redshift far-infrared luminous QSO has
a $^{12}$CO(9--8) luminosity L$_{\rm
  CO}\lsim$\,1\,$\times$\,10$^8$\,L$_{\odot}$.  Moreover, if the line
high-$J$ $^{12}$CO (presumably therefore a galaxy with an AGN), then
the gas temperature must be high ($\gsim$\,300\,K if the $^{12}$CO
spectral line energy distribution peaks beyond $J$\,=\,7).  Yet the
characteristic dust temperature would have to be $\lsim$\,15--25\,K
(for a dust emissivity of $\beta$\,=\,1.5--2.0).  In addition, if these
SMGs are lower-redshift ($z\sim$\,1--3) starbursts or AGN, then these
SMGs also lie significantly off the far-infrared--radio correlation
\citep{Condon91,Ivison10FAR-radio}.

Taken together, we suggest that the line is most likely a fine
structure transition, and given the brightness and equivalent width,
probably [C{\sc ii}].  We note that although the redshifts of these two
sources are similar ($\Delta v\sim$\,1500\,km\,s$^{-1}$), they are
separated by 25\,arcminutes on the sky (physical distance of 10\,Mpc)
and so they are not physically associated.


\subsection{Broad Band Spectral Energy Distribution}

\begin{figure}
  \centerline{
    \psfig{file=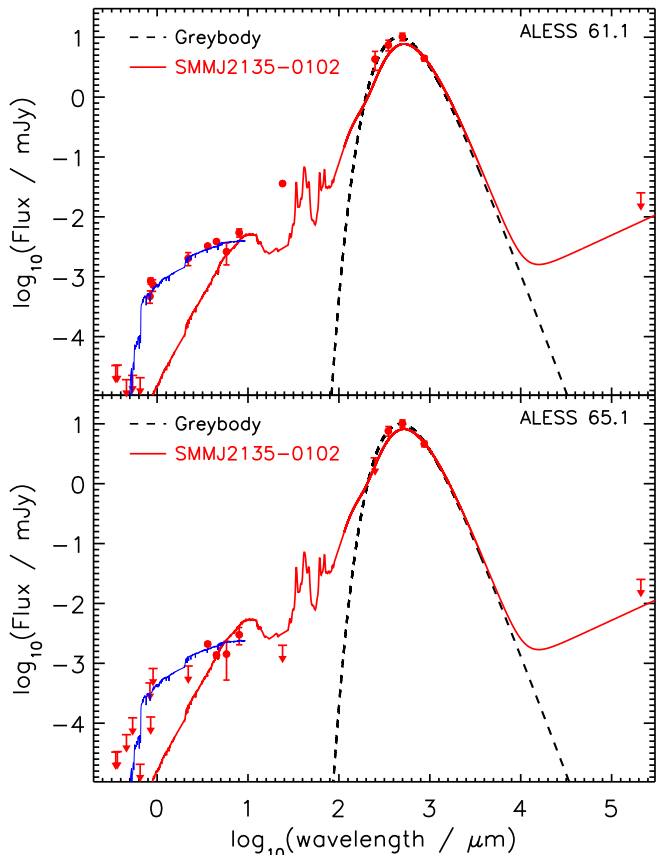,angle=0,width=3.3in}}
\caption{The observed optical--radio spectral energy distributions for
  ALESS\,61.1 and ALESS\,65.1.  We adopt redshifts for the two galaxies of
  $z$\,=\,4.4189 and $z$\,=\,4.4445.  We model the far-infrared SEDs
  using a single component modified black body dust model (dashed
  line) by fitting the 250-, 350-, 500-$\mu$m limits and 870-$\mu$m
  photometry.  We also model the rest-frame UV--near-infrared
  photometry using {\sc hyper-z} (blue line).  The red solid curve
  show the SED of the well studied $z$\,=\,2.3 starburst galaxy,
  SMM\,J2135, at the redshift of our galaxies (normalised to the
  870-$\mu$m flux density).  This demonstrates that the far-infrared
  colours are consistent with the high-redshift nature of these
  galaxies.}
\label{fig:SEDs}
\end{figure}

We show in Fig.~\ref{fig:SEDs} the observed broad-band SED for both
galaxies.  We overlay the SED of the well-studied $z$\,=\,2.3
starburst SMM\,J2135$-$0102 \citep{Swinbank10Nature,Ivison10eyelash}
redshifted to $z$\,=\,4.4 and normalised to the 870$\mu$m photometry.
This demonstrates that at $z\sim$\,4.4, the 250, 350 and 500$\mu$m
colours and 1.4\,GHz of these galaxies are consistent with the
expected colours of $z>$\,4 ULIRGs (assuming the properties of ULIRGs
at $z>$\,4 are the same as those at $z\sim$\,2;
\citealt{Schinnerer09,Coppin09,Daddi09,Capak08}).

We fit a modified black-body to the far-infrared photometry, deriving
estimates of the far-infrared (rest-frame 8--1000$\mu$m) luminosities
of $L_{\rm FIR}$\,=\,(2.1\,$\pm$\,0.4)\,$\times$10$^{12}$\,L$_{\odot}$
and $L_{\rm
  FIR}$\,=\,(2.0\,$\pm$\,0.4)\,$\times$10$^{12}$\,L$_{\odot}$, for
ALESS\,61.1 and ALESS\,65.1 respectively (corresponding to star-formation
rates of $\sim$\,500\,M$_{\odot}$\,yr$^{-1}$; \citealt{Kennicutt98}).


To estimate the near-infrared luminosities of these systems, we use the
rest-frame UV--mid-infrared SEDs.  We fit elliptical, Sb, single burst
and constant star-formation rate spectral templates from
\citet{Bruzual03} to the 0.3--8-$\mu$m photometry (Table~1 and
Fig.~\ref{fig:SEDs}) using {\sc hyper-z} \citep{Bolzonella00} at the
known redshift.  We allow reddening of A$_{\rm V}=$\,0--5 in steps of
0.2 and over-plot the best-fitting SED models in Fig.~\ref{fig:SEDs}.
We then calculate the rest-frame $H$-band magnitude (observed 8\,$\mu$m
at $z\sim$\,4.4), which is less influenced by young stars than
rest-frame UV or optical bands and is relatively unaffected by dust.
We derive absolute rest-frame $H$-band magnitudes of $H_{\rm
  AB}$\,=\,$-$24.8\,$\pm$\,0.2 and $H_{\rm AB}$\,=\,$-$24.2\,$\pm$\,0.3
for ALESS\,61.1 and ALESS\,65.1, respectively (Table~2).  These are
comparable to the average $H$-band magnitudes for radio-identified
sub-millimeter source counterparts at $z\lsim$\,3 ($H_{\rm
  AB}$\,=\,$-$24.1\,$\pm$\,0.9; \citealt{Wardlow11}), the bulk of which
are expected to be SMGs (Smail et al.\ 2012, in prep).

Owing to the catastrophic degeneracies between star-formation history,
age and reddenning in SED fitting for dusty sources, to estimate the
stellar masses we adot a simple approach.  Following
\citet{Hainline09} and \citet{Wardlow11} we use the $H$-band magnitude
together with an average mass-to-light ratio for a likely SMG
star-formation history.  \citet{Hainline09} estimate a $H$-band
mass-to-light ratio for SMGs (with burst and constant star-formation
templates), deriving an average $L_{\rm
  H}$\,/\,$M_{\star}\sim$\,3.8\,$L_{\odot}$\,/\,$M_{\odot}$ (for a
Salpeter IMF).  This suggests an stellar mass for ALESS\,61.1 and
ALESS\,65.1 of $M_{\star}\sim$\,1.5\,$\times$\,10$^{11}$ and
9\,$\times$\,10$^{10}$\,M$_{\odot}$ respectively, comparable to
previous estimates for SMG stellar masses
\citep[e.g.\ ][]{Hainline09,Wardlow11}.  However, we caution that
\citet{Wardlow11} show that the uncertainties in the derived spectral
types and ages result in an estimated factor of $\sim$\,5\,$\times$
uncertainty in assumed mass-to-light ratios and thus stellar masses.

%
%

\subsection{The {\rm [C{\sc ii}]} properties}


The [C{\sc ii}] luminosities of our two SMGs, $L_{\rm
  [CII]}$\,=\,1.5--3.2\,$\times$\,10$^{9}$\,L$_{\odot}$, are comparable
to the luminosities of other high-redshift ULIRGs and QSOs
(e.g. \citealt{Maiolino09,Walter09,Wagg10,Ivison10eyelash,DeBreuck11};
Walter et al.\ 2012).
We can crudely estimate the molecular gas mass for these galaxies using
the average [C{\sc ii}]\,/\,$^{12}$CO(1--0) emission line ratio
($L_{\rm [CII]}$\,/\,$L_{\rm CO(1-0)}$\,=\,4400$\pm$1000) from a sample
of ten $z$\,=\,2--4 starbursts
\citep[e.g.][]{HD10,Ivison10eyelash,Stacey10,DeBreuck11}, although we
caution that there is significant scatter in the CO/[C{\sc ii}] luminosity
ratio (e.g.\ Fig.~5 of \citealt{Stacey10} shows that there is a range
of $L_{\rm [CII]}$\,/\,$L_{\rm CO(1-0)}$ of approximately 1\,dex).
Nevertheless, from this compilation of high-redshift observations, the
conversion between [C{\sc ii}] line luminosity and molecular gas mass
scales approximately as $M_{\rm gas}$\,=\,10$\pm$2\,($L_{\rm
  C[II]}$\,/\,$L_{\odot}$), suggesting a molecular gas mass for these
two SMGs of $M_{\rm gas}$\,=\,1--4\,$\times$\,10$^{10}$\,M$_{\odot}$.
This molecular gas mass is comparable to the median gas mass implied
for SMGs at $z$\,=\,2, $M_{\rm
  gas}$\,=\,5$\pm$1$\times$10$^{10}$\,M$_{\odot}$ \citep{Bothwell12}.
Of course, direct measurements of the low-$J$ CO emission are required
to provide a more reliable estimate of the gas mass in these systems.

\begin{figure*}
  \centerline{
    \psfig{file=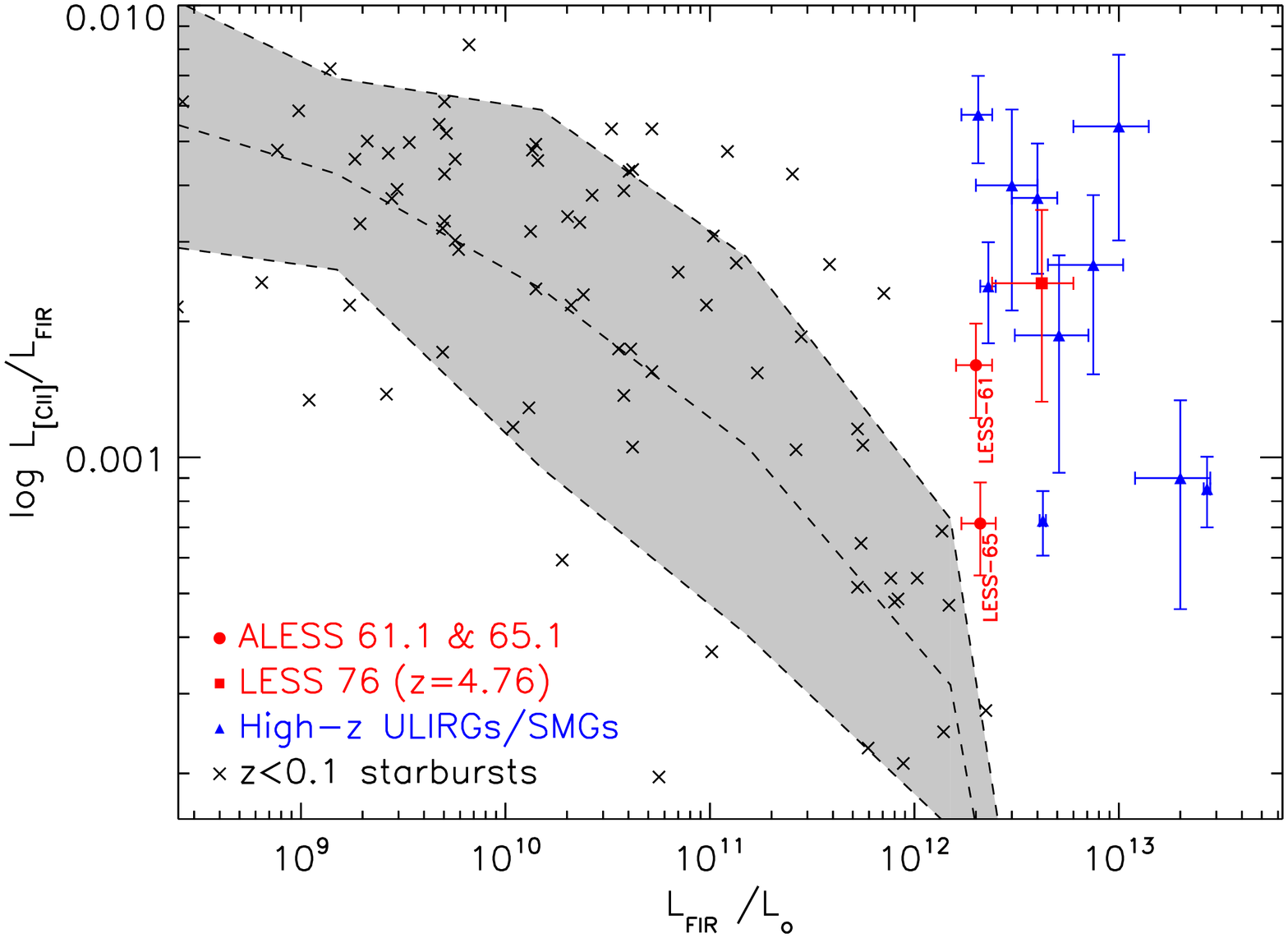,angle=90,width=3.6in}
    \psfig{file=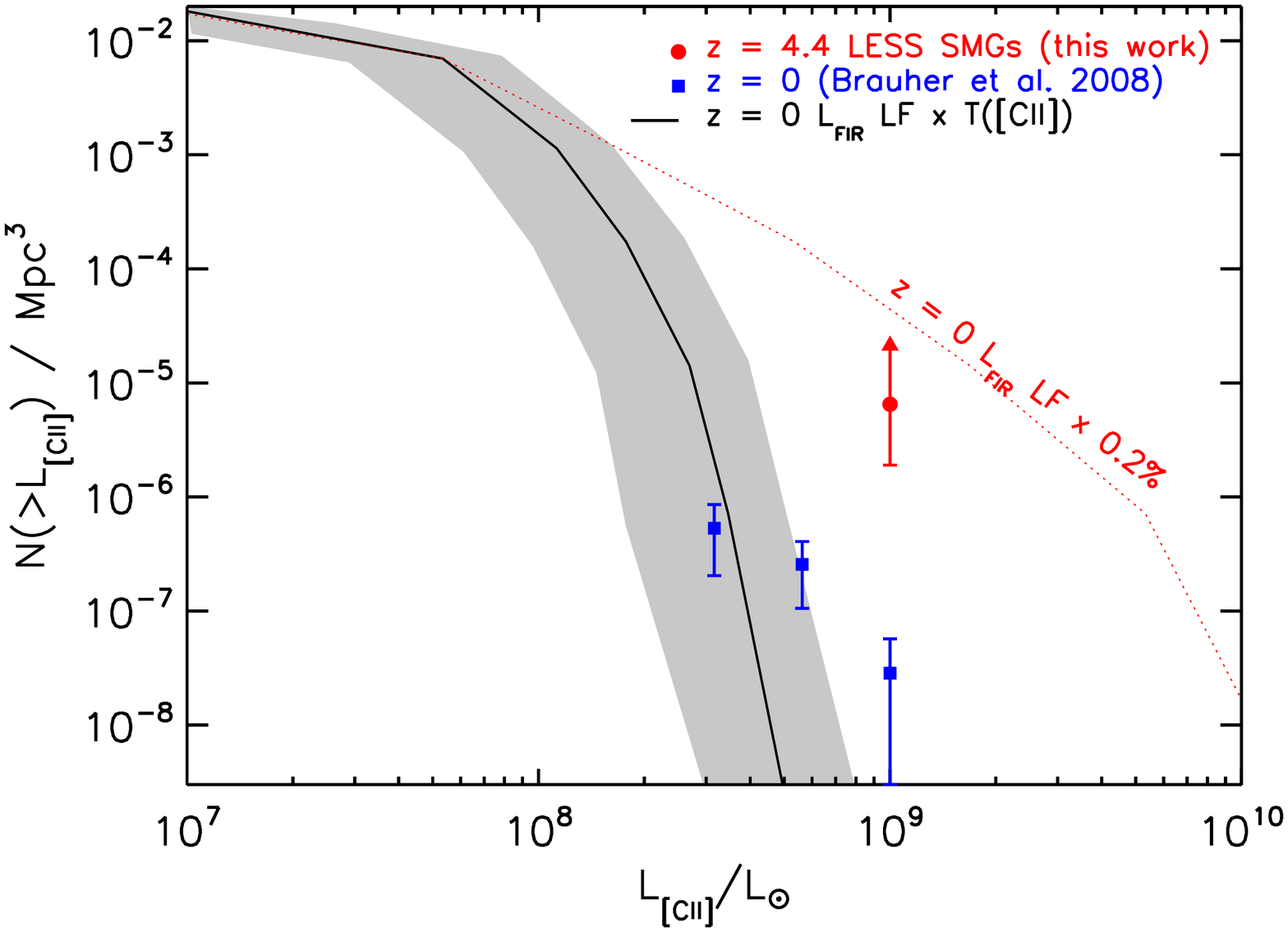,angle=90,width=3.55in}}
\caption{{\it Left:} L$_{\rm [CII]}$\,/\,L$_{\rm FIR}$ ratio as a
  function of the far-infrared luminosity for our two $z$\,=\,4.4 ALMA
  SMGs compared to local star-forming galaxies and ULIRGs.  In the plot
  we also include the $z$\,=\,4.76 LESS SMG from \citet{DeBreuck11}
  (see also \citealt{Coppin09}).  We also include a number of
  high-redshift starbursts and AGN from previous studies
  \citep{Malhotra01,Colbert99,Unger00,Spinoglio05,Carral94,Brauher08,Luhman98,Luhman03,Stacey10,Cox11}.
  For the local data, we calculate the median (dashed line) and scatter
  (grey).  This figure shows that the ratio of $L_{\rm
    [CII]}$\,/\,$L_{\rm FIR}$ for high-redshift ULIRGs is a factor
  $\sim$\,10\,$\times$ higher given their far-infrared luminosities
  compared to those at $z\sim$\,0.  {\it Right:} The [C{\sc ii}]
  luminosity function at $z$\,=\,4.4 from our survey compared to
  $z$\,=\,0.  For the $z$\,=\,4.4 luminosity function, we assume that
  all of the [C{\sc ii}]-emitting galaxies in the $\Delta z$\,=\,0.12
  volume covered by our observations were detected and so we stress
  that these calculations yield only a lower-limit on the volume
  density of high-redshift [C{\sc ii}]-emitters.  The $z$\,=\,0
  observations are derived from volume density of \emph{IRAS} sources
  at $z$\,$<$\,0.05 from \citet{Brauher08}.  The red dotted line shows
  the predicted [C{\sc ii}] local luminosity function for a constant
  $L_{\rm [C II]}$\,/\,$L_{\rm FIR}$\,=\,\,0.002, whilst the solid line
  shows the $z$\,=\,0 far-infrared luminosity function convolved with
  the $L_{\rm [C II]}$\,/\,$L_{\rm FIR}$ function from
  \citet{Brauher08} (with the grey region denoting the errors in the
  volume density of local ULIRGs and scatter in the transfer function).
  Finding two galaxies in this volume probed by our observations
  indicates a $\gsim$\,1000 factor increase in the number density of
  luminous [C{\sc ii}] emitters from $z\sim $\,0--4.4, equivalent to a
  $\sim$\,3--4\,$\times$ [C{\sc ii}] luminosity evolution between
  $z$\,=\,0--4.4.}
\label{fig:CII_LF}
\end{figure*}

We can also combine the [C{\sc ii}] and far-infrared luminosities to
derive a ratio of $L_{\rm [C II]}$\,/\,$L_{\rm
  FIR}$\,=\,(7.1\,$\pm$\,1.6)$\times$\,10$^{-4}$ and
(16.0\,$\pm$\,3.7)\,$\times$\,10$^{-4}$ for LESS\,61.1 and LESS\,65.1
respectively.  As Fig.~\ref{fig:CII_LF} shows, in local luminous
star-forming galaxies the ratio of $L_{\rm [CII]}$\,/\,$L_{\rm FIR}$
declines by a factor $\sim$\,100 over three orders of magnitude in
far-infrared luminosity (with a factor $\sim$\,3\,$\times$ scatter at
any $L_{\rm FIR}$).  This ``[C{\sc ii}] deficit'' in local ULIRGs has a
number of possible explanations, including enhanced contribution from
continuum emission in dusty, high-ionisation regions
\citep{Luhman98,Luhman03}, high-ionisation effects in the dense
environments \citep{Abel09} or enhanced contributions to the infrared
luminosity from AGN \citep{Sargysan12}.

The ratio we derive for the two $z\sim$\,4.4 SMGs is a factor
$\sim$\,10\,$\times$ higher than expected for $z\sim$\,0 ULIRGs with
their far-infrared luminosity (Fig.~\ref{fig:CII_LF}), although the
high $L_{\rm [C II]}$\,/\,$L_{\rm FIR}$ is similar to other
high-redshift galaxies of comparable far-infrared luminosity
(e.g.\ \citealt{Stacey10,Cox11}; Walter et al.\ 2012).  This has been
interpreted as evidence that the molecular emission does not reside in
a single, compact region illuminated by an intense UV radiation field
(as is a good approximation for local ULIRGs), but rather that the
[C{\sc ii}] reservoir is more extended, with the high L$_{\rm
  [CII]}$\,/\,L$_{\rm FIR}$ ratio reflecting the lower density of this
extended medium \citep[e.g.\ ][]{Abel09,HD10,Gracia-Carpio11}.  This
interpretation is also consistent with high-resolution studies of SMGs,
which have shown that the dust emission and gas reservoirs can be
extended over several kilo-parsecs
\citep[e.g.][]{Biggs08,Younger08,Tacconi08,Ivison11EVLA,Swinbank11} and
studies of their mid-infrared colors and spectral properties
\citep{Menendez07,Hainline09}.  As we will show in \S\ref{sec:vel}, the
      [C{\sc ii}] emission in ALESS\,65.1 appears to show velocity
      structure, and we determine that the [C{\sc ii}] reservoir may be
      extended across $\sim$\,3\,kpc, consistent with this
      interpretation.

%
%

Finally, we note that since the [C{\sc ii}] transition is a primary
coolant within the PDRs, it provides a probe of the physical conditions
of the gas and interstellar radiation field
\citep[e.g.][]{HollenbachTielens99,Wolfire03}.  We can therefore use
the [C{\sc ii}] and far-infrared luminosity together with the PDR
models of \citet{Kaufman99} to place upper limits on the far-UV
radiation field strength and characteristic density of star-forming
regions within the ISM.  The [C{\sc ii}]/$L_{\rm FIR}$ ratio of our two
galaxies, $L_{\rm [C II]}$\,/\,$L_{\rm
  FIR}$\,=\,7.1--16\,$\times$\,10$^{-4}$, suggests a mean interstellar
far-UV radiation field strength (G) $\lsim$\,3000$\times$ that of
Milky-Way and H$_2$ density $\lsim$\,10$^{5.5}$\,cm$^{-3}$
(e.g.\ \citealt{Danielson11}; see also \citealt{HD10} and
\citealt{Stacey10}).  The far-UV field strength limit derived in this
way is comparable to that estimated using {\sc starburst99}, which
suggests that a 500\,M$_{\odot}$\,yr$^{-1}$ starburst produces a
luminosity of $\sim$\,5\,$\times$\,10$^{45}$\,erg\,s$^{-1}$ at a
wavelength of 1000\,\AA\ or a flux density of $\sim$\,10$^{3.5}\times$
that of the Milky Way for a source with physical extent of 3\,kpc
(adopting a far-UV field strength between 912--1103\AA\ for the Milky
Way of 1.6\,$\times$\,10$^{-3}$\,erg\,s$^{-1}$\,cm$^{-2}$).

\begin{table}
\begin{center}
\caption{Physical Properties of the Galaxies}
\begin{tabular}{lll}
\hline
\hline
                       & ~~~ALESS\,61.1                                &  ~~~ALESS\,65.1                           \\
\hline
$\lambda_{\rm c}$       & 350.726\,$\pm$\,0.034\,$\mu$m               & 349.073\,$\pm$\,0.032\,$\mu$m                    \\
$z_{\rm [CII]}$         & 4.4189\,$\pm$\,0.0004                        & 4.4445\,$\pm$\,0.0005                     \\
$I_{\rm [CII]}$         & 2.5\,$\pm$\,0.4\,Jy\,km\,s$^{-1}$            & 5.4\,$\pm$\,0.7\,Jy\,km\,s$^{-1}$           \\
FWHM                  & 230\,$\pm$\,25\,km\,s$^{-1}$                 & 470$\pm$\,35\,km\,s$^{-1}$                 \\
$L_{\rm [CII]}$         & (1.5\,$\pm$\,0.3)\,$\times$\,10$^{9}$\,L$_{\odot}$  & (3.2\,$\pm$\,0.4)\,$\times$\,10$^{9}$\,L$_{\odot}$  \\
$L_{\rm FIR}$          & (2.1\,$\pm$\,0.4)\,$\times$\,10$^{12}$\,L$_{\odot}$ & (2.0$\pm$\,0.4)\,$\times$\,10$^{12}$\,L$_{\odot}$ \\
$H_{\rm AB}$           & $-$24.8\,$\pm$\,0.2                           & $-$24.2\,$\pm$\,0.3                            \\
$L_{\rm H}$            & (5.9\,$\pm$\,1.0)\,$\times$\,10$^{11}$\,L$_{\odot}$  & (3.4\,$\pm$\,0.8)\,$\times$\,10$^{11}$\,L$_{\odot}$ \\
$M_{\star}$            & 1.5$\times$\,10$^{11}$\,M$_{\odot}$               & 9.0$\times$\,10$^{10}$\,M$_{\odot}$  \\
\hline
\label{table:table1}
\end{tabular}
\end{center}
\noindent{\footnotesize Notes: The ALMA data have been primary beam
  corrected before calculating luminosities.  The primary beam
  corrections are 1.12 and 1.04 for ALESS\,61.1 and ALESS\,65.1
  respectively).  \hfil }
\end{table}

\subsection{Internal structure}
\label{sec:vel}

Given the spatial resolution of our observations ($\sim$1.4$''$ FWHM),
we search for spatially resolved emission and velocity structure within
these two SMGs.  From the ALMA data of LESS\,65, the total flux of the
galaxy ($S_{\rm 870}$\,=\,4.24\,$\pm$\,0.49\,mJy) is marginally higher
than the peak flux in the map ($S_{\rm
  peak}$\,=\,3.60\,$\pm$\,0.49\,mJy\,beam$^{-1}$) indicating that the
source may be marginally resolved.  Using the [C{\sc ii}] emission in
this galaxy, we examine whether there is any evidence for velocity
structure by comparing the spatial distribution of the integrated
[C{\sc ii}] emission between $-$210--0\,km\,s$^{-1}$ and
0--210\,km\,s$^{-1}$ (i.e. $\pm$FWHM relative to the systemic redshift
in the continuum subtracted cube).  By centroiding the two maps we
derive a spatial offset of 3.3\,$\pm$\,1.7\,kpc between the redshifted
and blue-shifted emission (Fig.~\ref{fig:maps}).  Although tentative,
this spatial extent is similar to the [C{\sc ii}] spatial extent of the
bright $z\sim$\,5 SMG HDF\,850.1 where similar data are available
\citep{Walter12}, and also to the high-$J$ $^{12}$CO molecular emission
FWHM ($\sim$4\,kpc) from a sample of eight SMGs at $z\sim2$
\citep{Tacconi06}.  If the spatial offset in ALESS\,65.1 represents
rotating gas, then we estimate a mass M$_{\rm
  dyn}\sim$\,3.5\,$\times$\,10$^{10}$\,sin$^2(i)$, which is consistent
with the implied gas and stellar mass estimates.




\subsection{The evolution of the {\rm [C{\sc ii}]} luminosity function}

Since our ALMA survey has followed up {\it all} of the LESS SMGs, we
can use the detection rate to investigate the evolution of the [C{\sc
    ii}] luminosity function.  In the following we assume that all of
the [C{\sc ii}]-emitting galaxies in the $\Delta z$\,=\,0.12 volume
covered by our observations were detected and so we stress that these
calculations yield only a lower-limit on the volume density of
high-redshift [C{\sc ii}]-emitters.

To estimate the $z\sim $\,0 [C{\sc ii}] luminosity function, we start
from the $z\sim $\,0 far-infrared luminosity function of
\citet{Sanders03} and initially assume a constant $L_{\rm
  [CII]}$\,/\,$L_{\rm FIR}$\,=\,0.002 (Fig.~\ref{fig:CII_LF}).  This
provides a firm upper limit to the bright end of $z\sim$\,0 [C{\sc ii}]
luminosity function, given the observed decline in $L_{\rm
  [CII]}$\,/\,$L_{\rm FIR}$ at high $L_{\rm FIR}$
(Fig.~\ref{fig:CII_LF}).  We can then use the correlation of $L_{\rm
  [CII]}$\,/\,$L_{\rm FIR}$ with $L_{\rm FIR}$ from \citet{Brauher08}
(see also Fig.~\ref{fig:CII_LF}) to provide a more reliable estimate.
We show our best estimate of the $z\sim$\,0 [C{\sc ii}] luminosity
function in Fig.~\ref{fig:CII_LF}, including the expected dispersion
due to the scatter in the local $L_{\rm [CII]}$\,/\,$L_{\rm
  FIR}$--$L_{\rm FIR}$ relation.

To assess the reliability of this $z\sim$\,0 [C{\sc ii}] luminosity
function we attempt to independently derive this using the
\citet{Brauher08} study of [C{\sc ii}] emission in a sample of 227
$z<$\,0.05 galaxies from the \emph{ISO} archive.  The parent population
of this study can be approximated by an \emph{IRAS} 100-$\mu$m selected
sample at $z<$\,0.05 with flux densities between S$_{\rm 100\mu
  m}=$\,1--1000\,Jy.  There are $\sim$\,11,000 {\it IRAS} sources
within this redshift and flux density range and so we must account for
the incompleteness in the \citet{Brauher08} sample to derive the $z
\sim$\,0 [C{\sc ii}] luminosity function.  We start by assuming the
\citet{Brauher08} sample is a random subset of the parent population
(although we note it is likely that apparently bright sources are
over-represented in the sample, and fainter sources correspondingly
under-represented).  We then only consider {\it IRAS} sources with
flux densities $>$\,10\,Jy where the fraction of galaxies observed in the
\citet{Brauher08} sample is $\gsim$\,3\% of the parent population and
then calculate the fraction of galaxies in the \citet{Brauher08} sample
compared to the number of {\it IRAS} galaxies at $z<$\,0.05 in bins of
100-$\mu$m flux density and redshift.  We use these sampling fractions
to correct the apparent [C{\sc ii}] luminosity function within
$z<$\,0.05 (a comoving volume of 0.035\,Gpc$^{3}$) and show these data
in Fig.~\ref{fig:CII_LF}.

Whilst the local data from \citet{Brauher08} comprise a complex mix of
observations derived from a number of studies, the data are broadly
consistent with the bright end of the $z\sim$\,0 [C{\sc ii}] luminosity
function we derived above from the far-infrared luminosity function and
$L_{\rm [CII]}$\,/\,$L_{\rm FIR}$ transfer function.

To place the new ALMA-identified $z$\,=\,4.4 SMGs on this plot, we
estimate the volume covered by our observations of ECDFS by considering
that the brightest [C{\sc ii}] emitters will correspond to the
brightest far-infrared sources (i.e.\ SMGs).  The LESS SMGs are located
within a 0.5$^{\circ}\times$\,0.5$^{\circ}$ region and assuming our
ALMA observations cover the [C{\sc ii}] with a redshift range of
$\Delta z$\,=\,0.12 at $z\sim$\,4.4, we derive a comoving volume of
2.9\,$\times$\,10$^{5}$\,Mpc$^3$.  Finding two galaxies in this volume
with [C{\sc ii}] luminosities $>$\,1\,$\times$\,10$^{9}$\,L$_{\odot}$
indicates a $\gsim$\,1000 factor increase in the number density of
luminous [C{\sc ii}] emitters from $z\sim $\,0--4.4.  This is
equivalent to a $\sim$\,3--4\,$\times$ [C{\sc ii}] luminosity evolution
between $z$\,=\,0--4.4 which is consistent with the evolution in the
far-infrared luminosity function (factor $\sim$\,3$\times$ in $L_{\rm
  FIR}$ for a fixed volume density of $\phi$\,=\,10$^{-5}$\,Mpc$^{-1}$;
\citealt{Wardlow11}) between $z$\,=\,0 and $z$\,=\,3, although since
the far-infrared luminosity function declines beyond $z$\,=\,3, the
[C{\sc ii}] luminosity function may peak at lower redshift than probed
by our observations here.

\section{Summary}

We have undertaken an ALMA study of 126 sub-millimetre sources from the
LABOCA 870$\mu$m survey of the ECDFS \citep{Weiss09}.  We focus here on
two high-redshift SMGs that are precisely located by our high
resolution ALMA continuum observations and whose ALMA spectra detect
bright emission lines.  We interpret these lines as the far-infrared
atomic fine structure line [C{\sc ii}]\,157.74, indicating redshifts of
$z$\,=\,4.4 for both galaxies.  We show that the ratio of $L_{\rm [C
    II]}$\,/\,$L_{\rm FIR}$ is higher than comparably luminous galaxies
at $z$\,=\,0, but consistent with the ratio seen in other high-redshift
ULIRGs.  We interpret this as evidence that the molecular emission is
extended and indeed, in one of our SMGs our data suggest that the
[C{\sc ii}] is resolved on $\gsim$\,3\,kpc scales.  Given the volume
probed by our observations, we show that the [C{\sc ii}] luminosity
function must evolve strongly across the 12\,Gyrs between $z$\,=\,0 and
$z\sim$\,4.

These results show that the wide spectral baseline coverage of ALMA
provides the opportunity to measure blind redshifts of large samples of
distant, obscured galaxies through the detection of fine structure
emission lines, such as [C{\sc ii}], in short exposure times.

\section*{acknowledgments}

We would like to thank the anonymous referee for a thoughtful and
constructive report which improved the content and clarity of this
paper.  The ALMA observations were carried out under program
2011.0.00294.S.  ALMA is a partnership of ESO (representing its member
states), NSF (USA) and NINS (Japan), together with NRC (Canada) and NSC
and ASIAA (Taiwan), in cooperation with the Republic of Chile. The
Joint ALMA Observatory is operated by ESO, AUI/NRAO and NAOJ.  This
publication is also based on data acquired with the APEX under
programme IDs 078.F-9028(A), 079.F-9500(A), 080.A-3023(A) and
081.F-9500(A). APEX is a collaboration between the Max-Planck-Institut
fur Radioastronomie, the European Southern Observatory and the Onsala
Space Observatory.  This research also made use of data from the HerMES
Key Programme from the SPIRE instrument team, ESAC scientists and a
mission scientist.  Herschel is an ESA space observatory with science
instruments provided by European-led Principal Investigator consortia
and with important participation from NASA.  AMS gratefully
acknowledges an STFC Advanced Fellowship.  IRS acknowledges support
from STFC and a Leverhume Fellowship.  KEKC acknowledges support from
the endowment of the Lorne Trottier Chair in Astrophysics and Cosmology
at McGill, the Natural Science and Engineering Research Council of
Canada (NSERC), and a L'Or\'{e}al Canada for Women in Science Research
Excellence Fellowship, with the support of the Canadian Commission for
UNESCO.  TRG acknowledges the Science and Technologies Facilities
Council as well as IDA and DARK.


\begin{thebibliography}{78}
\expandafter\ifx\csname natexlab\endcsname\relax\def\natexlab#1{#1}\fi

\bibitem[{{Abel} {et~al.}(2009){Abel}, {Dudley}, {Fischer}, {Satyapal}, \& {van
  Hoof}}]{Abel09}
{Abel}, N.~P., {Dudley}, C., {Fischer}, J., {Satyapal}, S., \& {van Hoof},
  P.~A.~M. 2009, \apj, 701, 1147

\bibitem[{{Biggs} \& {Ivison}(2008)}]{Biggs08}
{Biggs}, A.~D. \& {Ivison}, R.~J. 2008, \mnras, 385, 893

\bibitem[{{Biggs} {et~al.}(2011){Biggs}, {Ivison}, {Ibar}, {Wardlow},
  {Dannerbauer}, {Smail}, {Walter}, \& {Wei{\ss}, A. et al.}}]{Biggs11}
{Biggs}, A.~D., {Ivison}, R.~J., {Ibar}, E., {Wardlow}, J.~L., {Dannerbauer},
  H., {Smail}, I., {Walter}, F., \& {Wei{\ss}, A. et al.} 2011, \mnras, 413,
  2314

\bibitem[{{Blain} {et~al.}(1999){Blain}, {Smail}, {Ivison}, \&
  {Kneib}}]{Blain99b}
{Blain}, A.~W., {Smail}, I., {Ivison}, R.~J., \& {Kneib}, J.-P. 1999, \mnras,
  302, 632

\bibitem[{{Blain} {et~al.}(2002){Blain}, {Smail}, {Ivison}, {Kneib}, \&
  {Frayer}}]{Blain02}
{Blain}, A.~W., {Smail}, I., {Ivison}, R.~J., {Kneib}, J.-P., \& {Frayer},
  D.~T. 2002, PhR, 369, 111

\bibitem[{{Bolzonella} {et~al.}(2000){Bolzonella}, {Miralles}, \& {Pell{\'
  o}}}]{Bolzonella00}
{Bolzonella}, M., {Miralles}, J.-M., \& {Pell{\' o}}, R. 2000, \aap, 363, 476

\bibitem[{{Bothwell} {et~al.}(2012){Bothwell}, {Smail}, {Chapman}, {Genzel},
  {Ivison}, {Tacconi}, {Alaghband-Zadeh}, \& {Bertoldi, F. et
  al.}}]{Bothwell12}
{Bothwell}, M.~S., {Smail}, I., {Chapman}, S.~C., {Genzel}, R., {Ivison},
  R.~J., {Tacconi}, L.~J., {Alaghband-Zadeh}, S., \& {Bertoldi, F. et al.}
  2012, ArXiv e-prints

\bibitem[{{Brauher} {et~al.}(2008){Brauher}, {Dale}, \& {Helou}}]{Brauher08}
{Brauher}, J.~R., {Dale}, D.~A., \& {Helou}, G. 2008, \apjs, 178, 280

\bibitem[{{Bruzual} \& {Charlot}(2003)}]{Bruzual03}
{Bruzual}, G. \& {Charlot}, S. 2003, \mnras, 344, 1000

\bibitem[{{Capak} {et~al.}(2008){Capak}, {Carilli}, {Lee}, {Aldcroft},
  {Aussel}, {Schinnerer}, {Wilson}, \& {Yun, M.~S. et al.}}]{Capak08}
{Capak}, P., {Carilli}, C.~L., {Lee}, N., {Aldcroft}, T., {Aussel}, H.,
  {Schinnerer}, E., {Wilson}, G.~W., \& {Yun, M.~S. et al.} 2008, \apjl, 681,
  L53

\bibitem[{{Carilli} \& {Yun}(2000)}]{CarilliYun00}
{Carilli}, C.~L. \& {Yun}, M.~S. 2000, \apj, 530, 618

\bibitem[{{Carral} {et~al.}(1994){Carral}, {Hollenbach}, {Lord}, {Colgan},
  {Haas}, {Rubin}, \& {Erickson}}]{Carral94}
{Carral}, P., {Hollenbach}, D.~J., {Lord}, S.~D., {Colgan}, S.~W.~J., {Haas},
  M.~R., {Rubin}, R.~H., \& {Erickson}, E.~F. 1994, \apj, 423, 223

\bibitem[{{Chapman} {et~al.}(2005){Chapman}, {Blain}, {Smail}, \&
  {Ivison}}]{Chapman05a}
{Chapman}, S.~C., {Blain}, A.~W., {Smail}, I., \& {Ivison}, R.~J. 2005, \apj,
  622, 772

\bibitem[{{Chapman} {et~al.}(2004){Chapman}, {Smail}, {Windhorst}, {Muxlow}, \&
  {Ivison}}]{Chapman04b}
{Chapman}, S.~C., {Smail}, I., {Windhorst}, R., {Muxlow}, T., \& {Ivison},
  R.~J. 2004, \apj, 611, 732

\bibitem[{{Colbert} {et~al.}(1999){Colbert}, {Malkan}, {Clegg}, {Cox},
  {Fischer}, {Lord}, {Luhman}, \& {Satyapal, S. et al.}}]{Colbert99}
{Colbert}, J.~W., {Malkan}, M.~A., {Clegg}, P.~E., {Cox}, P., {Fischer}, J.,
  {Lord}, S.~D., {Luhman}, M., \& {Satyapal, S. et al.} 1999, \apj, 511, 721

\bibitem[{{Condon} {et~al.}(1991){Condon}, {Anderson}, \& {Helou}}]{Condon91}
{Condon}, J.~J., {Anderson}, M.~L., \& {Helou}, G. 1991, \apj, 376, 95

\bibitem[{{Coppin} {et~al.}(2009){Coppin}, {Smail}, {Alexander}, {Weiss},
  {Walter}, {Swinbank}, {Greve}, {Kovacs}, \& {De Breuck, C. et
  al.}}]{Coppin09}
{Coppin}, K.~E.~K., {Smail}, I., {Alexander}, D.~M., {Weiss}, A., {Walter}, F.,
  {Swinbank}, A.~M., {Greve}, T.~R., {Kovacs}, A., \& {De Breuck, C. et al.}
  2009, \mnras, 395, 1905

\bibitem[{{Cox} {et~al.}(2011){Cox}, {Krips}, {Neri}, {Omont}, {G{\"u}sten},
  {Menten}, {Wyrowski}, \& {Wei{\ss}, A. et al.}}]{Cox11}
{Cox}, P., {Krips}, M., {Neri}, R., {Omont}, A., {G{\"u}sten}, R., {Menten},
  K.~M., {Wyrowski}, F., \& {Wei{\ss}, A. et al.} 2011, \apj, 740, 63

\bibitem[{{Daddi} {et~al.}(2009){Daddi}, {Dannerbauer}, {Krips}, {Walter},
  {Dickinson}, {Elbaz}, \& {Morrison}}]{Daddi09}
{Daddi}, E., {Dannerbauer}, H., {Krips}, M., {Walter}, F., {Dickinson}, M.,
  {Elbaz}, D., \& {Morrison}, G.~E. 2009, \apjl, 695, L176

\bibitem[{{Danielson} {et~al.}(2011){Danielson}, {Swinbank}, {Smail}, {Cox},
  {Edge}, {Weiss}, {Harris}, \& {Baker, , A.~J. et al.}}]{Danielson11}
{Danielson}, A.~L.~R., {Swinbank}, A.~M., {Smail}, I., {Cox}, P., {Edge},
  A.~C., {Weiss}, A., {Harris}, A.~I., \& {Baker, , A.~J. et al.} 2011, \mnras,
  410, 1687

\bibitem[{{Dannerbauer} {et~al.}(2002){Dannerbauer}, {Lehnert}, {Lutz},
  {Tacconi}, {Bertoldi}, {Carilli}, {Genzel}, \& {Menten}}]{Dannerbauer02}
{Dannerbauer}, H., {Lehnert}, M.~D., {Lutz}, D., {Tacconi}, L., {Bertoldi}, F.,
  {Carilli}, C., {Genzel}, R., \& {Menten}, K. 2002, \apj, 573, 473

\bibitem[{{De Breuck} {et~al.}(2011){De Breuck}, {Maiolino}, {Caselli},
  {Coppin}, {Hailey-Dunsheath}, \& {Nagao}}]{DeBreuck11}
{De Breuck}, C., {Maiolino}, R., {Caselli}, P., {Coppin}, K.,
  {Hailey-Dunsheath}, S., \& {Nagao}, T. 2011, \aap, 530, L8

\bibitem[{{Decarli} {et~al.}(2012){Decarli}, {Walter}, {Neri}, {Bertoldi},
  {Carilli}, {Cox}, {Kneib}, \& {Lestrade, J.~F. et al.}}]{Decarli12}
{Decarli}, R., {Walter}, F., {Neri}, R., {Bertoldi}, F., {Carilli}, C., {Cox},
  P., {Kneib}, J.~P., \& {Lestrade, J.~F. et al.} 2012, \apj, 752, 2

\bibitem[{{Frayer} {et~al.}(2000){Frayer}, {Smail}, {Ivison}, \&
  {Scoville}}]{Frayer00}
{Frayer}, D.~T., {Smail}, I., {Ivison}, R.~J., \& {Scoville}, N.~Z. 2000, \aj,
  120, 1668

\bibitem[{{Gear} {et~al.}(2000){Gear}, {Lilly}, {Stevens}, {Clements}, {Webb},
  {Eales}, \& {Dunne}}]{Gear00}
{Gear}, W.~K., {Lilly}, S.~J., {Stevens}, J.~A., {Clements}, D.~L., {Webb},
  T.~M., {Eales}, S.~A., \& {Dunne}, L. 2000, \mnras, 316, L51

\bibitem[{{Graci{\'a}-Carpio} {et~al.}(2011){Graci{\'a}-Carpio}, {Sturm},
  {Hailey-Dunsheath}, {Fischer}, {Contursi}, {Poglitsch}, {Genzel}, \&
  et~al.}]{Gracia-Carpio11}
{Graci{\'a}-Carpio}, J., {Sturm}, E., {Hailey-Dunsheath}, S., {Fischer}, J.,
  {Contursi}, A., {Poglitsch}, A., {Genzel}, R., \& et~al., G. 2011, \apjl,
  728, L7

\bibitem[{{Hailey-Dunsheath} {et~al.}(2010){Hailey-Dunsheath}, {Nikola},
  {Stacey}, {Oberst}, {Parshley}, {Benford}, {Staguhn}, \& {Tucker}}]{HD10}
{Hailey-Dunsheath}, S., {Nikola}, T., {Stacey}, G.~J., {Oberst}, T.~E.,
  {Parshley}, S.~C., {Benford}, D.~J., {Staguhn}, J.~G., \& {Tucker}, C.~E.
  2010, \apjl, 714, L162

\bibitem[{{Hailey-Dunsheath} {et~al.}(2008){Hailey-Dunsheath}, {Nikola},
  {Stacey}, {Oberst}, {Parshley}, {Bradford}, {Ade}, \& {Tucker}}]{SHD08}
{Hailey-Dunsheath}, S., {Nikola}, T., {Stacey}, G.~J., {Oberst}, T.~E.,
  {Parshley}, S.~C., {Bradford}, C.~M., {Ade}, P.~A.~R., \& {Tucker}, C.~E.
  2008, \apjl, 689, L109

\bibitem[{{Hainline} {et~al.}(2009){Hainline}, {Blain}, {Smail}, {Frayer},
  {Chapman}, {Ivison}, \& {Alexander}}]{Hainline09}
{Hainline}, L.~J., {Blain}, A.~W., {Smail}, I., {Frayer}, D.~T., {Chapman},
  S.~C., {Ivison}, R.~J., \& {Alexander}, D.~M. 2009, \apj, 699, 1610

\bibitem[{{Hollenbach} \& {Tielens}(1999)}]{HollenbachTielens99}
{Hollenbach}, D.~J. \& {Tielens}, A.~G.~G.~M. 1999, Reviews of Modern Physics,
  71, 173

\bibitem[{{Ivison} {et~al.}(2007){Ivison}, {Greve}, {Dunlop}, {Peacock},
  {Egami}, {Smail}, \& {Ibar et al.}}]{Ivison07}
{Ivison}, R.~J., {Greve}, T.~R., {Dunlop}, J.~S., {Peacock}, J.~A., {Egami},
  E., {Smail}, I., \& {Ibar et al.} 2007, \mnras, 380, 199

\bibitem[{{Ivison} {et~al.}(2002){Ivison}, {Greve}, {Smail}, {Dunlop}, {Roche},
  {Scott}, {Page}, \& {Stevens et al.}}]{Ivison02}
{Ivison}, R.~J., {Greve}, T.~R., {Smail}, I., {Dunlop}, J.~S., {Roche}, N.~D.,
  {Scott}, S.~E., {Page}, M.~J., \& {Stevens et al.} 2002, \mnras, 337, 1

\bibitem[{{Ivison} {et~al.}(1998){Ivison}, {Harrison}, \&
  {Coulson}}]{Ivison98b}
{Ivison}, R.~J., {Harrison}, A.~P., \& {Coulson}, I.~M. 1998, \aap, 330, 443

\bibitem[{{Ivison} {et~al.}(2010{\natexlab{a}}){Ivison}, {Magnelli}, {Ibar},
  {Andreani}, {Elbaz}, {Altieri}, {Amblard}, \& {Arumugam, V. et
  al.}}]{Ivison10FAR-radio}
{Ivison}, R.~J., {Magnelli}, B., {Ibar}, E., {Andreani}, P., {Elbaz}, D.,
  {Altieri}, B., {Amblard}, A., \& {Arumugam, V. et al.} 2010{\natexlab{a}},
  \aap, 518, L31

\bibitem[{{Ivison} {et~al.}(2011){Ivison}, {Papadopoulos}, {Smail}, {Greve},
  {Thomson}, {Xilouris}, \& {Chapman}}]{Ivison11EVLA}
{Ivison}, R.~J., {Papadopoulos}, P.~P., {Smail}, I., {Greve}, T.~R., {Thomson},
  A.~P., {Xilouris}, E.~M., \& {Chapman}, S.~C. 2011, \mnras, 412, 1913

\bibitem[{{Ivison} {et~al.}(2010{\natexlab{b}}){Ivison}, {Swinbank},
  {Swinyard}, {Smail}, {Pearson}, {Rigopoulou}, {Polehampton}, \& {Baluteau,
  J.-P.. et al.}}]{Ivison10eyelash}
{Ivison}, R.~J., {Swinbank}, A.~M., {Swinyard}, B., {Smail}, I., {Pearson},
  C.~P., {Rigopoulou}, D., {Polehampton}, E., \& {Baluteau, J.-P.. et al.}
  2010{\natexlab{b}}, \aap, 518, L35

\bibitem[{{Kaufman} {et~al.}(1999){Kaufman}, {Wolfire}, {Hollenbach}, \&
  {Luhman}}]{Kaufman99}
{Kaufman}, M.~J., {Wolfire}, M.~G., {Hollenbach}, D.~J., \& {Luhman}, M.~L.
  1999, \apj, 527, 795

\bibitem[{{Kennicutt}(1998)}]{Kennicutt98}
{Kennicutt}, R.~C. 1998, \araa, 36, 189

\bibitem[{{LeFloc'h} {et~al.}(2009){LeFloc'h}, {Aussel}, {Ilbert}, {Riguccini},
  {Frayer}, {Salvato}, {Arnouts}, \& {Surace, J. et al.}}]{LeFloch09}
{LeFloc'h}, E., {Aussel}, H., {Ilbert}, O., {Riguccini}, L., {Frayer}, D.~T.,
  {Salvato}, M., {Arnouts}, S., \& {Surace, J. et al.} 2009, \apj, 703, 222

\bibitem[{{Lindner} {et~al.}(2011){Lindner}, {Baker}, {Omont}, {Beelen},
  {Owen}, {Bertoldi}, {Dole}, \& {Fiolet, N. et al.}}]{Lindner11}
{Lindner}, R.~R., {Baker}, A.~J., {Omont}, A., {Beelen}, A., {Owen}, F.~N.,
  {Bertoldi}, F., {Dole}, H., \& {Fiolet, N. et al.} 2011, \apj, 737, 83

\bibitem[{{Lord} {et~al.}(1996){Lord}, {Malhotra}, {Lim}, {Helou}, {Rubin},
  {Stacey}, {Hollenbach}, \& {Werner, M.~W. et al}}]{Lord96}
{Lord}, S.~D., {Malhotra}, S., {Lim}, T., {Helou}, G., {Rubin}, R.~H.,
  {Stacey}, G.~J., {Hollenbach}, D.~J., \& {Werner, M.~W. et al}. 1996, \aap,
  315, L117

\bibitem[{{Luhman} {et~al.}(1998){Luhman}, {Satyapal}, {Fischer}, {Wolfire},
  {Cox}, {Lord}, {Smith}, {Stacey}, \& {Unger}}]{Luhman98}
{Luhman}, M.~L., {Satyapal}, S., {Fischer}, J., {Wolfire}, M.~G., {Cox}, P.,
  {Lord}, S.~D., {Smith}, H.~A., {Stacey}, G.~J., \& {Unger}, S.~J. 1998,
  \apjl, 504, L11

\bibitem[{{Luhman} {et~al.}(2003){Luhman}, {Satyapal}, {Fischer}, {Wolfire},
  {Sturm}, {Dudley}, {Lutz}, \& {Genzel}}]{Luhman03}
{Luhman}, M.~L., {Satyapal}, S., {Fischer}, J., {Wolfire}, M.~G., {Sturm}, E.,
  {Dudley}, C.~C., {Lutz}, D., \& {Genzel}, R. 2003, \apj, 594, 758

\bibitem[{{Madden} {et~al.}(1993){Madden}, {Geis}, {Genzel}, {Herrmann},
  {Jackson}, {Poglitsch}, {Stacey}, \& {Townes}}]{Madden93}
{Madden}, S.~C., {Geis}, N., {Genzel}, R., {Herrmann}, F., {Jackson}, J.,
  {Poglitsch}, A., {Stacey}, G.~J., \& {Townes}, C.~H. 1993, \apj, 407, 579

\bibitem[{{Maiolino} {et~al.}(2009){Maiolino}, {Caselli}, {Nagao}, {Walmsley},
  {De Breuck}, \& {Meneghetti}}]{Maiolino09}
{Maiolino}, R., {Caselli}, P., {Nagao}, T., {Walmsley}, M., {De Breuck}, C., \&
  {Meneghetti}, M. 2009, \aap, 500, L1

\bibitem[{{Maiolino} {et~al.}(2005){Maiolino}, {Cox}, {Caselli}, {Beelen},
  {Bertoldi}, {Carilli}, {Kaufman}, \& {Menten, K.~M. et al.}}]{Maiolino05}
{Maiolino}, R., {Cox}, P., {Caselli}, P., {Beelen}, A., {Bertoldi}, F.,
  {Carilli}, C.~L., {Kaufman}, M.~J., \& {Menten, K.~M. et al.} 2005, \aap,
  440, L51

\bibitem[{{Malhotra} {et~al.}(2001){Malhotra}, {Kaufman}, {Hollenbach},
  {Helou}, {Rubin}, {Brauher}, {Dale}, \& {Lu, N.~Y. et al.}}]{Malhotra01}
{Malhotra}, S., {Kaufman}, M.~J., {Hollenbach}, D., {Helou}, G., {Rubin},
  R.~H., {Brauher}, J., {Dale}, D., \& {Lu, N.~Y. et al.} 2001, \apj, 561, 766

\bibitem[{{McMullin} {et~al.}(2007){McMullin}, {Waters}, {Schiebel}, {Young},
  \& {Golap}}]{McMullin07}
{McMullin}, J.~P., {Waters}, B., {Schiebel}, D., {Young}, W., \& {Golap}, K.
  2007, in Astronomical Society of the Pacific Conference Series, Vol. 376,
  Astronomical Data Analysis Software and Systems XVI, ed. R.~A. {Shaw},
  F.~{Hill}, \& D.~J. {Bell}, 127

\bibitem[{{Men{\'e}ndez-Delmestre} {et~al.}(2007){Men{\'e}ndez-Delmestre},
  {Blain}, {Alexander}, {Smail}, {Armus}, {Chapman}, {Frayer}, {Ivison}, \&
  {Teplitz}}]{Menendez07}
{Men{\'e}ndez-Delmestre}, K., {Blain}, A.~W., {Alexander}, D.~M., {Smail}, I.,
  {Armus}, L., {Chapman}, S.~C., {Frayer}, D.~T., {Ivison}, R.~J., \&
  {Teplitz}, H.~I. 2007, \apjl, 655, L65

\bibitem[{{Nagao} {et~al.}(2012){Nagao}, {Maiolino}, {De Breuck}, {Caselli},
  {Hatsukade}, \& {Saigo}}]{Nagao12}
{Nagao}, T., {Maiolino}, R., {De Breuck}, C., {Caselli}, P., {Hatsukade}, B.,
  \& {Saigo}, K. 2012, eprint arXiv:1205.4834

\bibitem[{{Pope} {et~al.}(2006){Pope}, {Scott}, {Dickinson}, {Chary},
  {Morrison}, {Borys}, {Sajina}, \& {Alexander, D. et al.}}]{Pope06}
{Pope}, A., {Scott}, D., {Dickinson}, M., {Chary}, R.-R., {Morrison}, G.,
  {Borys}, C., {Sajina}, A., \& {Alexander, D. et al.} 2006, \mnras, 370, 1185

\bibitem[{{Sakamoto} {et~al.}(2008){Sakamoto}, {Wang}, {Wiedner}, {Wang},
  {Peck}, {Zhang}, {Petitpas}, \& {Ho, L., et al.}}]{Sakamoto08}
{Sakamoto}, K., {Wang}, J., {Wiedner}, M.~C., {Wang}, Z., {Peck}, A.~B.,
  {Zhang}, Q., {Petitpas}, G.~R., \& {Ho, L., et al.} 2008, \apj, 684, 957

\bibitem[{{Sanders} {et~al.}(2003){Sanders}, {Mazzarella}, {Kim}, {Surace}, \&
  {Soifer}}]{Sanders03}
{Sanders}, D.~B., {Mazzarella}, J.~M., {Kim}, D.-C., {Surace}, J.~A., \&
  {Soifer}, B.~T. 2003, \aj, 126, 1607

\bibitem[{{Sanders} \& {Mirabel}(1996)}]{Sanders96}
{Sanders}, D.~B. \& {Mirabel}, I.~F. 1996, \araa, 34, 749

\bibitem[{{Sargsyan} {et~al.}(2012){Sargsyan}, {Lebouteiller}, {Weedman},
  {Spoon}, {Bernard-Salas}, {Engels}, {Stacey}, {Houck}, {Barry}, {Miles}, \&
  {Samsonyan}}]{Sargysan12}
{Sargsyan}, L., {Lebouteiller}, V., {Weedman}, D., {Spoon}, H.,
  {Bernard-Salas}, J., {Engels}, D., {Stacey}, G., {Houck}, J., {Barry}, D.,
  {Miles}, J., \& {Samsonyan}, A. 2012, ArXiv e-prints

\bibitem[{{Schinnerer} {et~al.}(2008){Schinnerer}, {Carilli}, {Capak},
  {Martinez-Sansigre}, {Scoville}, {Smol{\v c}i{\'c}}, {Taniguchi}, {Yun},
  {Bertoldi}, {Le Fevre}, \& {de Ravel}}]{Schinnerer09}
{Schinnerer}, E., {Carilli}, C.~L., {Capak}, P., {Martinez-Sansigre}, A.,
  {Scoville}, N.~Z., {Smol{\v c}i{\'c}}, V., {Taniguchi}, Y., {Yun}, M.~S.,
  {Bertoldi}, F., {Le Fevre}, O., \& {de Ravel}, L. 2008, \apjl, 689, L5

\bibitem[{{Smail} {et~al.}(2011){Smail}, {Swinbank}, {Ivison}, \&
  {Ibar}}]{Smail11}
{Smail}, I., {Swinbank}, A.~M., {Ivison}, R.~J., \& {Ibar}, E. 2011, \mnras,
  414, L95

\bibitem[{{Smol{\v c}i{\'c}} {et~al.}(2012){Smol{\v c}i{\'c}}, {Navarrete},
  {Aravena}, {Ilbert}, {Yun}, {Sheth}, {Salvato}, \& {McCracken, H.~J. et
  al.}}]{Smolcic12}
{Smol{\v c}i{\'c}}, V., {Navarrete}, F., {Aravena}, M., {Ilbert}, O., {Yun},
  M.~S., {Sheth}, K., {Salvato}, M., \& {McCracken, H.~J. et al.} 2012, \apjs,
  200, 10

\bibitem[{{Spinoglio} {et~al.}(2005){Spinoglio}, {Malkan}, {Smith},
  {Gonz{\'a}lez-Alfonso}, \& {Fischer}}]{Spinoglio05}
{Spinoglio}, L., {Malkan}, M.~A., {Smith}, H.~A., {Gonz{\'a}lez-Alfonso}, E.,
  \& {Fischer}, J. 2005, \apj, 623, 123

\bibitem[{{Stacey} {et~al.}(1991){Stacey}, {Geis}, {Genzel}, {Lugten},
  {Poglitsch}, {Sternberg}, \& {Townes}}]{Stacey91}
{Stacey}, G.~J., {Geis}, N., {Genzel}, R., {Lugten}, J.~B., {Poglitsch}, A.,
  {Sternberg}, A., \& {Townes}, C.~H. 1991, \apj, 373, 423

\bibitem[{{Stacey} {et~al.}(2010){Stacey}, {Hailey-Dunsheath}, {Ferkinhoff},
  {Nikola}, {Parshley}, {Benford}, {Staguhn}, \& {Fiolet}}]{Stacey10}
{Stacey}, G.~J., {Hailey-Dunsheath}, S., {Ferkinhoff}, C., {Nikola}, T.,
  {Parshley}, S.~C., {Benford}, D.~J., {Staguhn}, J.~G., \& {Fiolet}, N. 2010,
  \apj, 724, 957

\bibitem[{{Swinbank} {et~al.}(2011){Swinbank}, {Papadopoulos}, {Cox}, {Krips},
  {Ivison}, {Smail}, {Thomson}, {Neri}, {Richard}, \& {Ebeling}}]{Swinbank11}
{Swinbank}, A.~M., {Papadopoulos}, P.~P., {Cox}, P., {Krips}, M., {Ivison},
  R.~J., {Smail}, I., {Thomson}, A.~P., {Neri}, R., {Richard}, J., \&
  {Ebeling}, H. 2011, \apj, 742, 11

\bibitem[{{Swinbank} {et~al.}(2010){Swinbank}, {Smail}, {Longmore}, {Harris},
  {Baker}, {De Breuck}, {Richard}, \& {Edge, A.~C. et al.}}]{Swinbank10Nature}
{Swinbank}, A.~M., {Smail}, I., {Longmore}, S., {Harris}, A.~I., {Baker},
  A.~J., {De Breuck}, C., {Richard}, J., \& {Edge, A.~C. et al.} 2010, \nat,
  464, 733

\bibitem[{{Tacconi} {et~al.}(2008){Tacconi}, {Genzel}, {Smail}, {Neri},
  {Chapman}, {Ivison}, {Blain}, \& {Cox, P., et al.}}]{Tacconi08}
{Tacconi}, L.~J., {Genzel}, R., {Smail}, I., {Neri}, R., {Chapman}, S.~C.,
  {Ivison}, R.~J., {Blain}, A., \& {Cox, P., et al.} 2008, \apj, 680, 246

\bibitem[{{Tacconi} {et~al.}(2006){Tacconi}, {Neri}, {Chapman}, {Genzel},
  {Smail}, {Ivison}, {Bertoldi}, \& {Blain et al.}}]{Tacconi06}
{Tacconi}, L.~J., {Neri}, R., {Chapman}, S.~C., {Genzel}, R., {Smail}, I.,
  {Ivison}, R.~J., {Bertoldi}, F., \& {Blain et al.} 2006, \apj, 640, 228

\bibitem[{{Unger} {et~al.}(2000){Unger}, {Clegg}, {Stacey}, {Cox}, {Fischer},
  {Greenhouse}, {Lord}, \& {Luhman, M.~L. et al.}}]{Unger00}
{Unger}, S.~J., {Clegg}, P.~E., {Stacey}, G.~J., {Cox}, P., {Fischer}, J.,
  {Greenhouse}, M., {Lord}, S.~D., \& {Luhman, M.~L. et al.} 2000, \aap, 355,
  885

\bibitem[{{Valtchanov} {et~al.}(2011){Valtchanov}, {Virdee}, {Ivison},
  {Swinyard}, {van der Werf}, {Rigopoulou}, {da Cunha}, \& {Lupu R. et
  al.}}]{Valtchanov11}
{Valtchanov}, I., {Virdee}, J., {Ivison}, R.~J., {Swinyard}, B., {van der
  Werf}, P., {Rigopoulou}, D., {da Cunha}, E., \& {Lupu R. et al.} 2011,
  \mnras, 415, 3473

\bibitem[{{van der Werf} {et~al.}(2010){van der Werf}, {Isaak}, {Meijerink},
  {Spaans}, {Rykala}, {Fulton}, {Loenen}, \& {Walter, F. et
  al.}}]{vanderWerf10}
{van der Werf}, P.~P., {Isaak}, K.~G., {Meijerink}, R., {Spaans}, M., {Rykala},
  A., {Fulton}, T., {Loenen}, A.~F., \& {Walter, F. et al.} 2010, \aap, 518,
  L42

\bibitem[{{Wagg} {et~al.}(2010){Wagg}, {Carilli}, {Wilner}, {Cox}, {De Breuck},
  {Menten}, {Riechers}, \& {Walter}}]{Wagg10}
{Wagg}, J., {Carilli}, C.~L., {Wilner}, D.~J., {Cox}, P., {De Breuck}, C.,
  {Menten}, K., {Riechers}, D.~A., \& {Walter}, F. 2010, \aap, 519, L1

\bibitem[{{Walter} {et~al.}(2012){Walter}, {Decarli}, {Carilli}, {Bertoldi},
  {Cox}, {da Cunha}, {Daddi}, \& {Dickinson, M. et al.}}]{Walter12}
{Walter}, F., {Decarli}, R., {Carilli}, C., {Bertoldi}, F., {Cox}, P., {da
  Cunha}, E., {Daddi}, E., \& {Dickinson, M. et al.} 2012, \nat, 486, 233

\bibitem[{{Walter} {et~al.}(2009{\natexlab{a}}){Walter}, {Riechers}, {Cox},
  {Neri}, {Carilli}, {Bertoldi}, {Weiss}, \& {Maiolino}}]{Walter09}
{Walter}, F., {Riechers}, D., {Cox}, P., {Neri}, R., {Carilli}, C., {Bertoldi},
  F., {Weiss}, A., \& {Maiolino}, R. 2009{\natexlab{a}}, \nat, 457, 699

\bibitem[{{Walter} {et~al.}(2009{\natexlab{b}}){Walter}, {Wei{\ss}},
  {Riechers}, {Carilli}, {Bertoldi}, {Cox}, \& {Menten}}]{Walter09b}
{Walter}, F., {Wei{\ss}}, A., {Riechers}, D.~A., {Carilli}, C.~L., {Bertoldi},
  F., {Cox}, P., \& {Menten}, K.~M. 2009{\natexlab{b}}, \apjl, 691, L1

\bibitem[{{Wang} {et~al.}(2011){Wang}, {Cowie}, {Barger}, \&
  {Williams}}]{Wang11}
{Wang}, W.-H., {Cowie}, L.~L., {Barger}, A.~J., \& {Williams}, J.~P. 2011,
  \apjl, 726, L18

\bibitem[{{Wardlow} {et~al.}(2011){Wardlow}, {Smail}, {Coppin}, {Alexander},
  {Brandt}, {Danielson}, {Luo}, \& {Swinbank, A.~M. et al.}}]{Wardlow11}
{Wardlow}, J.~L., {Smail}, I., {Coppin}, K.~E.~K., {Alexander}, D.~M.,
  {Brandt}, W.~N., {Danielson}, A.~L.~R., {Luo}, B., \& {Swinbank, A.~M. et
  al.} 2011, \mnras, 415, 1479

\bibitem[{{Wei{\ss}} {et~al.}(2009){Wei{\ss}}, {Kov{\'a}cs}, {Coppin}, {Greve},
  {Walter}, {Smail}, {Dunlop}, \& {Knudsen, K.~K. et al.}}]{Weiss09}
{Wei{\ss}}, A., {Kov{\'a}cs}, A., {Coppin}, K., {Greve}, T.~R., {Walter}, F.,
  {Smail}, I., {Dunlop}, J.~S., \& {Knudsen, K.~K. et al.} 2009, \apj, 707,
  1201

\bibitem[{{Wolfire} {et~al.}(2003){Wolfire}, {McKee}, {Hollenbach}, \&
  {Tielens}}]{Wolfire03}
{Wolfire}, M.~G., {McKee}, C.~F., {Hollenbach}, D., \& {Tielens}, A.~G.~G.~M.
  2003, \apj, 587, 278

\bibitem[{{Younger} {et~al.}(2007){Younger}, {Fazio}, {Huang}, {Yun}, {Wilson},
  {Ashby}, {Gurwell}, \& {Lai et al.}}]{Younger07}
{Younger}, J.~D., {Fazio}, G.~G., {Huang}, J.-S., {Yun}, M.~S., {Wilson},
  G.~W., {Ashby}, M.~L.~N., {Gurwell}, M.~A., \& {Lai et al.} 2007, \apj, 671,
  1531

\bibitem[{{Younger} {et~al.}(2008){Younger}, {Fazio}, {Wilner}, {Ashby},
  {Blundell}, {Gurwell}, {Huang}, \& {Iono, D. et al.}}]{Younger08}
{Younger}, J.~D., {Fazio}, G.~G., {Wilner}, D.~J., {Ashby}, M.~L.~N.,
  {Blundell}, R., {Gurwell}, M.~A., {Huang}, J.-S., \& {Iono, D. et al.} 2008,
  \apj, 688, 59

\end{thebibliography}

\end{document}